\documentclass[%
 reprint, 
superscriptaddress,
nofootinbib,
 amsmath,amssymb,
 aps, prd,
floatfix,
]{revtex4-2}

\usepackage{siunitx}
\usepackage{graphicx}
\usepackage{dcolumn}
\usepackage{bm}
\usepackage{multirow}

\usepackage[utf8]{inputenc} 
\usepackage[T1]{fontenc}    
\usepackage{xcolor}
\definecolor{darkblue}{RGB}{46,48,147}
\usepackage{url}            
\usepackage{booktabs}       
\usepackage{amsfonts}       
\usepackage{nicefrac}       
\usepackage{microtype}      
\usepackage{lipsum}
\usepackage{amsmath}
\usepackage{xspace}
\usepackage{array}
\usepackage{adjustbox}
\usepackage[acronym]{glossaries}
\usepackage{scalerel}
\usepackage{tikz}
\usetikzlibrary{svg.path}

\usepackage[colorlinks=true,
            linkcolor=darkblue,
            urlcolor=darkblue,
            citecolor=darkblue]{hyperref}     
\usepackage[capitalise]{cleveref}

\definecolor{orcidlogocol}{HTML}{A6CE39}
\tikzset{
  orcidlogo/.pic={
    \fill[orcidlogocol] svg{M256,128c0,70.7-57.3,128-128,128C57.3,256,0,198.7,0,128C0,57.3,57.3,0,128,0C198.7,0,256,57.3,256,128z};
    \fill[white] svg{M86.3,186.2H70.9V79.1h15.4v48.4V186.2z}
                 svg{M108.9,79.1h41.6c39.6,0,57,28.3,57,53.6c0,27.5-21.5,53.6-56.8,53.6h-41.8V79.1z M124.3,172.4h24.5c34.9,0,42.9-26.5,42.9-39.7c0-21.5-13.7-39.7-43.7-39.7h-23.7V172.4z}
                 svg{M88.7,56.8c0,5.5-4.5,10.1-10.1,10.1c-5.6,0-10.1-4.6-10.1-10.1c0-5.6,4.5-10.1,10.1-10.1C84.2,46.7,88.7,51.3,88.7,56.8z};
  }
}

\newcommand\orcidicon[1]{\href{https://orcid.org/#1}{\mbox{\scalerel*{
\begin{tikzpicture}[yscale=-1,transform shape]
\pic{orcidlogo};
\end{tikzpicture}
}{0}}}}

\newacronym{hpo}{HPO}{hyperparameter optimization}
\newacronym{hpc}{HPC}{high performance computing}
\newacronym{hep}{HEP}{high energy physics}
\newacronym{hp}{HP}{Hyperparameter}
\newacronym{ml}{ML}{machine learning}
\newacronym{nn}{NN}{neural network}
\newacronym{pf}{PF}{particle-flow}
\newacronym{oc}{OC}{object condensation}
\newacronym{gnn}{GNN}{graph neural network}
\newacronym{llm}{LLM}{large language model}
\newacronym{mlpf}{MLPF}{machine-learned particle flow}
\newacronym{iqr}{IQR}{interquartile range}
\newacronym{lhc}{LHC}{Large Hadron Collider}
\newacronym{hllhc}{HL-LHC}{High Luminosity LHC}
\newacronym{fcc}{FCC}{Future Circular Collider}
\newacronym{cms}{CMS}{Compact Muon Solenoid}
\newacronym{lsh}{LSH}{locality sensitive hashing}
\newacronym{cld}{CLD}{CLIC-like detector}
\newacronym{hgcal}{HGCAL}{high-granularity calorimeter}
\newacronym{knn}{kNN}{k-nearest neighbors}

\newcommand{\KVAL}{128}

\newcommand{\PW}{\textrm{W}\xspace}

\newcommand{\PGe}{\textrm{e}\xspace}

\newcommand{\PQs}{\textrm{s}\xspace}
\newcommand{\PQc}{\textrm{c}\xspace}
\newcommand{\PQb}{\textrm{b}\xspace}
\newcommand{\PQt}{\textrm{t}\xspace}

\newcommand{\PQAt}{\ensuremath{\overline{\PQt}}\xspace}

\newcommand{\ttbar}{\ensuremath{\PQt\PQAt}\xspace}

\newcommand{\PB}{\textrm{B}\xspace}
\newcommand{\PD}{\textrm{D}\xspace}

\newcommand{\pt}{\ensuremath{p_{\mathrm{T}}}\xspace}

\newcommand{\pythia} {{\textsc{Pythia}}\xspace}
\newcommand{\GEANTfour}{{\textsc{Geant4}}\xspace}
\newcommand{\ptmomentum}{\ensuremath{p_{\mathrm{T}}}\xspace}
\newcommand{\ptmiss}{\ensuremath{\ptmomentum^\text{miss}}\xspace}

\newcommand{\ptmissReco}{\ensuremath{\ptmomentum^\text{miss,reco}}\xspace}
\newcommand{\ptmissGen}{\ensuremath{\ptmomentum^\text{miss,gen}}\xspace}
\newcommand{\ptvecmiss}{\ensuremath{{\vec p}_{\mathrm{T}}^{\kern1pt\,\text{miss}}}\xspace}
\newcommand{\pvecmiss}{\ensuremath{{\vec p}^{\kern1pt\,\text{miss}}}\xspace}

\newcommand{\kt}{\ensuremath{k_{\mathrm{T}}}\xspace}

\newcommand{\GeV}{\ensuremath{\,\text{Ge\hspace{-.08em}V}}\xspace}

\newcommand{\baseline} {\textsc{Baseline}\xspace}
\newcommand{\latents} {\textsc{Latent-augmented}\xspace}
\newcommand{\probe} {\textsc{Linear-probe}\xspace}
\newcommand{\baselineL} {\textsc{Baseline-large}\xspace}
\newcommand{\probeL} {\textsc{Nonlinear-probe}\xspace}

\begin{document}

\title{\textbf{Machine-learned particle flow as a foundation model for collider physics} 
}%

\author{Farouk Mokhtar\orcidicon{0000-0003-2533-3402}}
 \email[Contact author: ]{fmokhtar@ucsd.edu}
 \thanks{\\(work done primarily while at UCSD)}
\affiliation{%
    \href{https://ror.org/0168r3w48}{University of California San Diego}, La Jolla, California 92093, USA
}%
\affiliation{%
    \href{https://ror.org/01ggx4157}{European Organization for Nuclear Research (CERN)}, Geneva 1211, Switzerland
}%
\author{Joosep Pata\orcidicon{0000-0002-5191-5759}}
\affiliation{
    \href{https://ror.org/03eqd4a41}{National Institute of Chemical Physics and Biophysics}, Tallinn 12618, Estonia
}%
\author{Michael Kagan\orcidicon{0000-0002-3386-6869}}
\affiliation{%
    \href{https://ror.org/05gzmn429}{SLAC National Accelerator Laboratory}, Menlo Park, California 94025, USA
}%
\author{Javier Duarte\orcidicon{0000-0002-5076-7096}}%
\affiliation{%
    \href{https://ror.org/0168r3w48}{University of California San Diego}, La Jolla, California 92093, USA
}%

\begin{abstract}

The workflow from particle collision to physics analysis passes through a series of reconstruction steps that are traditionally modular and disconnected, with no shared representation linking low-level detector data to high-level analysis tasks.
We show that casting event reconstruction as a machine learning problem naturally produces such a shared representation.
We repurpose a machine learning model trained for particle-flow reconstruction (MLPF) to perform three distinct analysis tasks: jet flavor identification, jet energy regression, and missing momentum regression.
By appending the per-particle latent representations learned during reconstruction as additional input features, we substantially improve over baselines that use conventional particle-level features alone.
We further demonstrate that a single linear layer trained using only the latent representations achieves competitive performance against state-of-the-art baseline architectures, and outperforms the baseline for missing momentum regression with approximately 35 times fewer parameters.
These results demonstrate that the latent representations learned during reconstruction encode essential physics information needed for downstream analysis, establishing MLPF as a foundation model and offering a concrete step toward an end-to-end pipeline from detector data to physics analysis.

\end{abstract}

\maketitle

\clearpage


\section{Introduction}
\label{sec:intro}

In collider experiments, the workflow from particle collision to physics analysis passes through a series of reconstruction steps that are traditionally modular and disconnected, with no shared representation linking low-level detector data to high-level analysis tasks.
Translating raw detector signals into physics objects, a task known as event reconstruction, is typically solved using the particle-flow (PF) algorithm~\cite{Sirunyan:2017ulk, Buskulic:1994wz}, which combines information from all sub-detectors to produce a list of stable particle candidates, or PF candidates.
This list then serves as the interface between the detector and all downstream physics analyses.
Once it is produced, however, the rich low-level correlations encoded in the detector signals are no longer directly accessible.
Recovering task-relevant information beyond the PF candidate four-momenta typically requires hand-engineering additional input features from lower-level quantities.
For instance, jet flavor identification relies on track-displacement variables that must be carefully derived from raw track parameters.

\begin{figure}[htpb]
  \centering
  \includegraphics[width=0.43\textwidth]{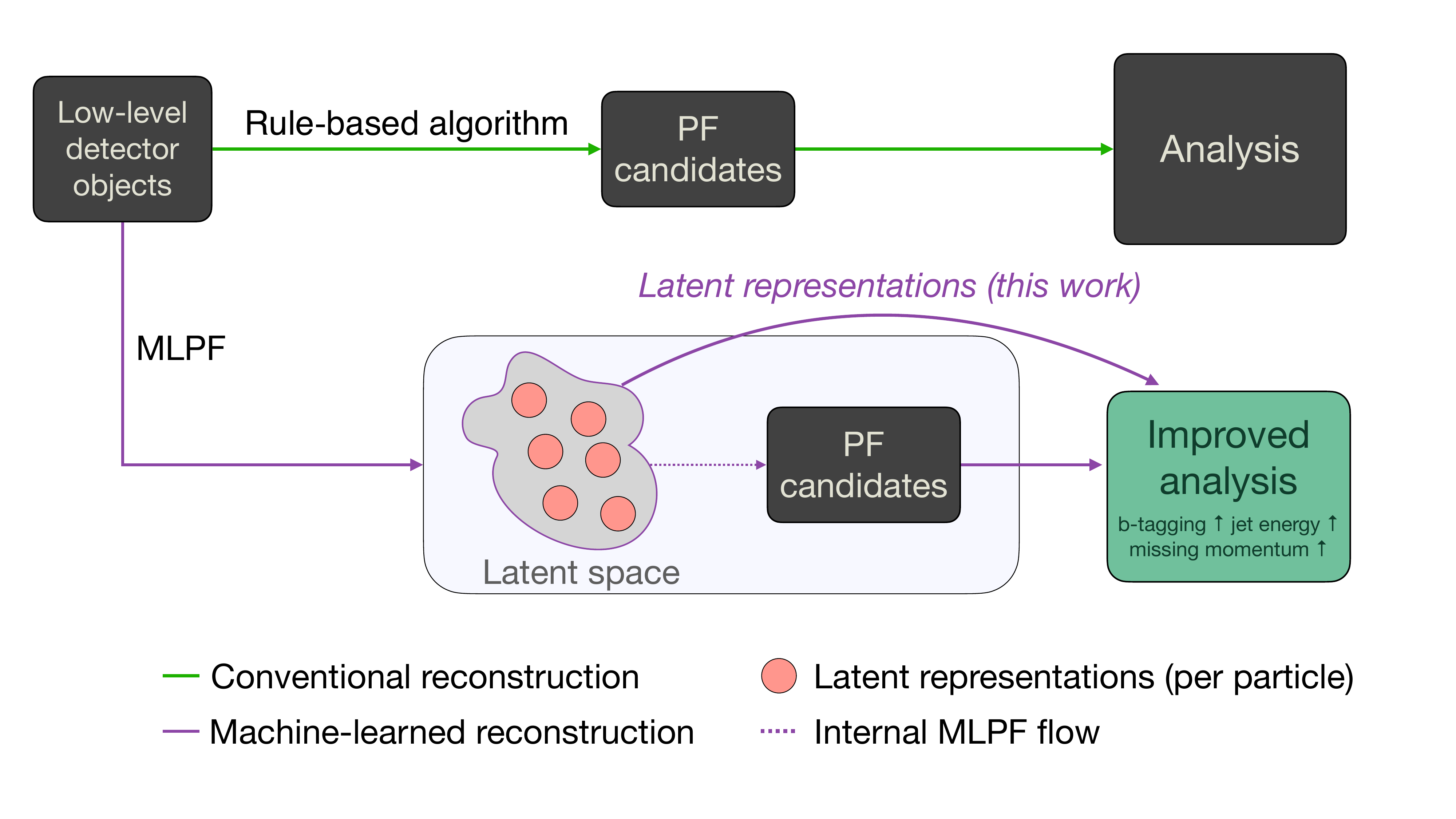}
  \caption{
  Schematic comparison of conventional and machine-learned particle-flow reconstruction.
  In the machine-learned approach, MLPF learns a rich per-particle latent representation that improves performance across different analysis tasks.
  }
  \label{fig:illustration}
\end{figure}

Machine learning (ML)-based reconstruction changes this picture.
Models such as machine-learned particle flow (MLPF)~\cite{Pata:2021oez} learn an internal high-dimensional representation of each collision event---referred to hereafter as \textit{latent representations}---as an intermediate step before reconstructing the PF candidates.
These representations are learned end-to-end and encode the full complexity of the detector response and particle interactions, including lower-level information that would otherwise require hand-engineering from raw quantities.
In this paper, we show that they also encode information useful beyond reconstruction.
By appending the per-particle latent representations to the standard PF candidate features, we substantially improve performance across three downstream analysis tasks that probe different levels of event complexity: jet flavor identification, which requires track-level quantities such as impact parameters (IPs) that lie below the PF candidate interface; jet energy regression, which operates at the jet constituent level; and missing momentum (\pvecmiss) regression, which aggregates information across the full event.
Figure~\ref{fig:illustration} contrasts the conventional pipeline, which produces only a list of PF candidates, with the ML-based pipeline, which additionally provides the latent representations to analysis tasks.

The prospect of building \emph{foundation models}, large models pretrained on broad data that can be adapted or fine-tuned to a wide range of downstream tasks~\cite{Bommasani2021}, for collider physics has attracted significant recent interest, with several works demonstrating that models pretrained on large jet datasets can be fine-tuned for downstream tasks with improved physics performance~\cite{Golling2023arXiv, Birk:2024knn, leigh2024token, harris2024resim, Kishimoto2023, Mikuni2024, Ho:2024arXiv, EveNet:2025,Amram:2024fjg, vigl2024finetuning, Li:2024htp, Zhao:2024kry, Mikuni:2025tar, Bhimji:2025isp}.
For instance, Ref.~\cite{vigl2024finetuning} demonstrates that jet-level latent representations from a pretrained transformer-based jet tagger can be input to an analysis-level event classifier to improve signal sensitivity significantly compared to using only conventional high-level features.
Reviews of this emerging field are given in Refs.~\cite{Hallin:2025fmh,Duarte:2025qbk}.

In parallel, MLPF reconstruction has become a viable alternative to rule-based PF, with approaches developed for multiple detector geometries~\cite{Pata:2023rhh, Mokhtar:2025mlpf,garcia2023towards,Wahlen:2024rxt,Garcia:2026reco} and recently commissioned in CMS data~\cite{CMS:MLPF}.
In Ref.~\cite{Mokhtar:2025mlpf}, we further demonstrated that an MLPF model pretrained on one detector geometry can be rapidly fine-tuned to a new detector with substantially less data than training from scratch, establishing cross-detector transfer as a first step toward a foundation model for collider physics.
   
The present work extends this picture by demonstrating cross-task transfer.
Rather than explicitly designing a foundation model from the outset, we demonstrate that MLPF already produces representations that are generically useful for downstream analysis tasks.
The MLPF backbone requires no modification, additional pretraining, or fine-tuning; its latent representations are extracted as a byproduct of standard reconstruction inference.
This suggests that ML-based reconstruction and physics analysis need not be treated as separate pipeline stages, and that reconstruction models may serve as natural foundation models for collider physics.

The paper is organized as follows.
Section~\ref{sec:mlpf} describes the MLPF architecture and extraction of the latent representations.
Section~\ref{sec:setup} presents the experimental setup.
Sections~\ref{sec:btag}, \ref{sec:jer}, and~\ref{sec:met} describe the three downstream tasks and report the results for each.
Section~\ref{sec:discussion} discusses the findings and their broader implications, and Section~\ref{sec:conclusion} summarizes the main conclusions.

\vspace{-10pt}

\section{MLPF latent representations}
\label{sec:mlpf}

The MLPF algorithm~\cite{Pata:2021oez} was originally designed as a graph neural network (GNN) and later evolved into a transformer-based architecture~\cite{Mokhtar:2025mlpf,CMS:MLPF}.
The precise underlying mechanism, whether message passing or attention, is not central to the following discussion.
In both cases, the model processes a set of input detector elements (tracks and calorimeter clusters) and learns to map them into a high-dimensional \textit{latent space}.
Unlike the input space, which is defined by raw detector measurements, the latent space is learned.
Its dimensions have no fixed physical meaning but are optimized during training to capture the correlations and structures in the data most useful for solving the reconstruction task.
The model then uses this representation to predict the kinematic properties (four-momentum) and particle identity (PID) of each reconstructed PF candidate.
Throughout this paper, we refer to the particles reconstructed by MLPF as PF candidates, using the same terminology as conventional PF algorithms to emphasize that MLPF serves as a drop-in replacement for traditional reconstruction.
  
In the latest implementation~\cite{Mokhtar:2025mlpf} used here, the latent space has 2048 dimensions.
This arises from a design choice in the MLPF architecture to maintain two separate 1024-dimensional encoder branches, one specialized for particle classification and the other for momentum regression.
This allows the classification and regression objectives to be optimized jointly during training while keeping their internal representations distinct.
This separation is not fundamental.
A single unified encoder of equivalent capacity would serve our purposes equally well.
We extract per-particle latent representations from the layer immediately preceding the output heads.
These representations encode structural information about the collision event that conventional reconstruction algorithms discard once the list of PF candidates is produced.

Since the full 2048-dimensional space is impractical as a direct input to downstream models, and to control model capacity, we apply a simple unsupervised compression step to retain the top $k$ dimensions (details in Sec.~\ref{sec:setup:pca}).
The specific choice of compression technique is not central to our argument.
We deliberately use a well-established, parameter-free method to demonstrate that the improvement is robust and does not depend on task-specific optimization of the compression.
We expect further gains from more sophisticated approaches, which we leave to future work.

\begin{figure}[hb]
\centering
\includegraphics[width=0.282\textwidth]{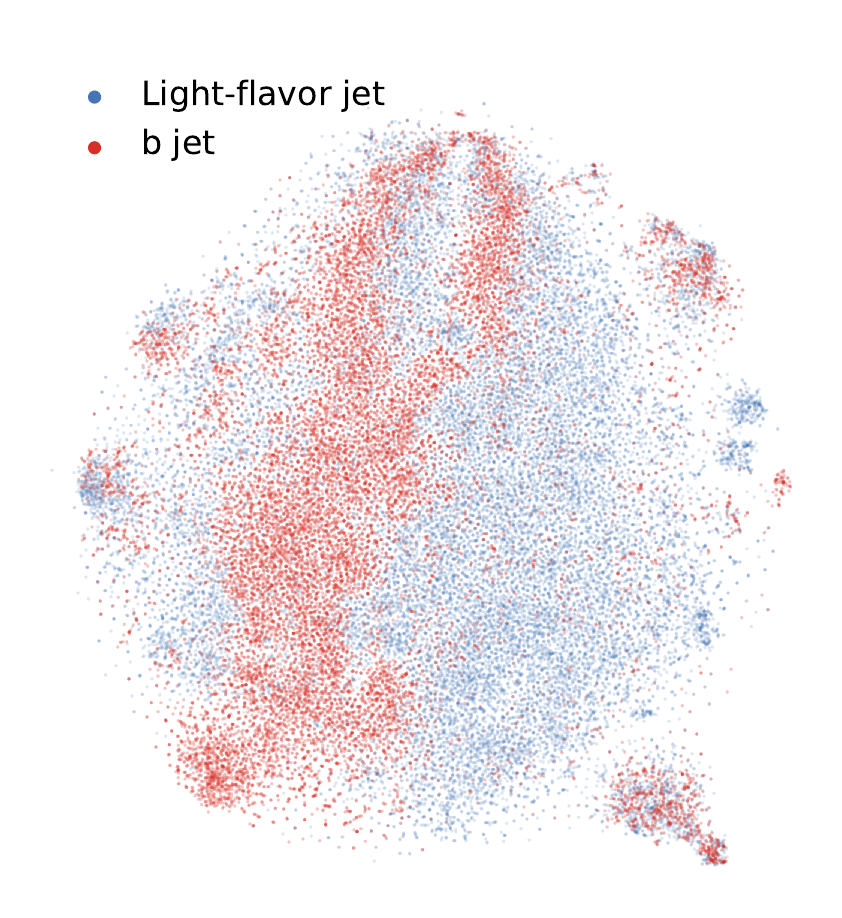}
\caption{
    Two-dimensional t-SNE projection of the MLPF latent representations averaged across the constituents of each jet on a sample of 50{,}000 jets, colored by generator-level flavor (red: \PQb jets; blue: light-flavor jets).
    The MLPF backbone is never trained using jet-flavor labels, yet the latent representations exhibit visible flavor separation.
}
\label{fig:tsne}
\end{figure}

As a qualitative illustration of the structure captured by the MLPF latent representations, Fig.~\ref{fig:tsne} shows a two-dimensional t-distributed stochastic neighbor embedding (t-SNE)~\cite{vanderMaaten:2008tsne} projection of latent representations averaged across the constituents of each jet in a sample of 50{,}000 jets.
Each point is colored by its generator-level flavor label (defined in Sec.~\ref{sec:btag}).
Despite the fact that the MLPF backbone never used jet-flavor labels during training, the projection shows visibly separated regions of \PQb jets and light-flavor jets, suggesting that jet-flavor information is encoded in the latent representations.

\section{Experimental setup}
\label{sec:setup}

Throughout this work, we use the MLPF model in Ref.~\cite{Mokhtar:2025mlpf} as our backbone model.
The MLPF backbone is pretrained on simulated datasets from the Compact Linear Collider detector (CLICdet) and subsequently fine-tuned on simulated datasets from the CLIC-like detector (CLD)~\cite{bacchetta2019cld}, a general-purpose detector concept proposed for the Future Circular Collider in electron-positron mode (FCC-ee)~\cite{FCC-ee}.
The model was trained using a total of 25 million events; more details in Ref.~\cite{Mokhtar:2025mlpf}.
This is the most recent MLPF checkpoint trained on the CLD geometry and is therefore the natural choice given that all events in this paper are simulated with the same detector geometry.

We adopt two deliberately conservative choices throughout.
First, the MLPF backbone weights remain completely frozen.
This approach is practically useful for collider analyses because analysts can leverage MLPF latent representations for downstream tasks without modifying the reconstruction pipeline or retraining the backbone.
Second, all downstream experiments use events drawn exclusively from the held-out test split of the MLPF fine-tuning procedure, so training-set contamination does not arise by construction and the observed improvements are attributable only to genuine transfer of learned representations.
Earlier studies indicated that allowing overlap with the MLPF fine-tuning set yields further performance gains, suggesting additional improvements are achievable when this constraint is relaxed.
We nevertheless adopt the stricter setup to remove any contamination concern.

The remainder of this section presents the dataset splits, the compression of the MLPF latent representations, and the evaluation design used in all three downstream studies.

\subsection{Dataset}
\label{sec:setup:dataset}

We use the CLD dataset described in Ref.~\cite{Mokhtar:2025mlpf}, consisting of top quark pair production ($\PGe^+\PGe^- \to \ttbar$) events at a center-of-mass energy of 365\GeV.
The CLD detector concept is a general-purpose detector sharing the same broad design philosophy as the ATLAS and CMS detectors at the LHC, comprising a silicon pixel and strip tracking system, a high-granularity electromagnetic calorimeter, a hadronic calorimeter, and a superconducting solenoid.
Events are generated with \pythia~\cite{Bierlich:2022pfr} and passed through full \GEANTfour~\cite{GEANT42003,GEANT42006,GEANT42016} simulation of the CLD geometry, providing a realistic description of particle interactions and detector response.
Each event contains a variable number of PF candidates, characterized by their transverse momentum (\pt), pseudorapidity ($\eta$), azimuthal angle ($\phi$), energy ($E$), and PID.
The all-inclusive \ttbar sample is well-suited for this study as it naturally provides all objects relevant to the downstream tasks considered: \PQb jets from top quark decays, \PQc jets from $\PW \to \PQc\overline{\PQs}$ decays, light-flavor jets from hadronic \PW boson decays, and genuine \pvecmiss from neutrinos produced in leptonic \PW boson decays.

The dataset is divided into three non-overlapping splits as shown in Table~\ref{tab:splits}.
An additional 5{,}000 events are reserved exclusively for the latent compression step described in the following subsection.

\begin{table}[htbp]
\centering
\begin{tabular}{l@{\hspace{1.5em}}r@{\hspace{1.5em}}l}
\toprule
Split & Events & Purpose \\
\midrule
Train      & 350{,}000 & Train downstream models \\
Validation &   5{,}000 & Track validation loss   \\
Test       &  40{,}000 & Benchmark physics performance   \\
\bottomrule
\end{tabular}
\caption{
Dataset splits used in this work.
All events are drawn from the MLPF held-out test split of Ref.~\cite{Mokhtar:2025mlpf}.
}
\label{tab:splits}
\end{table}

\subsection{PCA compression of latent representations}
\label{sec:setup:pca}
 
As described in Sec.~\ref{sec:mlpf}, the MLPF encoder produces a 2048-dimensional latent vector per PF candidate.
To reduce this to a computationally practical size, we apply principal component analysis (PCA) in an entirely unsupervised manner, using only the geometry of the MLPF latent space and introducing no task-specific information during the compression stage.
We intentionally use a task-agnostic compression to demonstrate that the MLPF latent representations carry generic information useful across different downstream tasks.

We apply PCA to the 2048-dimensional latent vectors, projecting them onto their $k$ leading principal components.
These are the $k$ directions that capture the most variance in the latent space.
We fit the projection on the dedicated 5{,}000 events reserved for this purpose, yielding approximately 744{,}000 PF candidate latent vectors, using the \texttt{IncrementalPCA} implementation from scikit-learn~\cite{sklearn} to avoid loading all vectors into memory at once.
We choose $k = 128$.
For this value, the projection retains 55\% of the total variance.
The cumulative explained variance as a function of $k$ is notably flat, reflecting a latent space where information is broadly distributed across many dimensions rather than concentrated in a few dominant directions.
The resulting $k$-dimensional compressed vectors are appended as additional per-particle features alongside the standard kinematic inputs.
During inference, the compression reduces to a single $2048 \times k$ matrix multiply per PF candidate (${\sim}0.3$\,M FLOPs), adding negligible overhead, orders of magnitude below the MLPF inference and downstream-model costs.

\subsection{Evaluation design}
\label{sec:setup:design}

For each downstream task, we train and evaluate three model variants:

\begin{itemize}
    
    \item \textbf{\baseline}: established task-specific architecture with standard PF candidate kinematic features as input (four-momentum, PID).

    \item \textbf{\latents}: same architecture and inputs as \baseline, augmented with the $\KVAL$-dimensional MLPF latent vector per PF candidate.

    \item \textbf{\probe}: a single linear layer using only the $\KVAL$-dimensional MLPF latent vector representations per PF candidate as input.

\end{itemize}

Particle identity is encoded as a one-hot vector over the five MLPF output classes (charged hadron, neutral hadron, photon, electron, muon).
Depending on the downstream task, additional input features beyond PF candidate kinematics and PID may be provided to the \baseline and \latents models.
For jet flavor identification, for instance, track IP variables are included.
In all cases we follow established conventions from prior publications, as detailed in Secs.~\ref{sec:btag}--\ref{sec:met}.

Since the \latents model receives additional input features, its parameter count is inherently larger than that of the \baseline model even with identical hidden layer widths.
To verify that the observed gains are not simply a consequence of this increased capacity, we also train a \baselineL model with the same inputs as \baseline but with wider hidden layers matching the parameter count of \latents.
We observe that the \baselineL model performs comparably to the \baseline model across all three tasks, confirming that capacity alone does not explain the performance gains.
Full results are reported in Appendix~\ref{app:capacity}.

The \probe model serves a distinct diagnostic purpose.
In representation learning, a \textit{linear probe}~\cite{alain2016understanding} is a single linear layer trained on top of a frozen embedding, quantifying how much task-relevant information is linearly accessible in the latent space, independently of any nonlinear processing by the downstream architecture.
A strong linear probe result indicates that the latent representations intrinsically organize the relevant physics signal in a geometrically separable way.
In the \probe model, all learned parameters reside in a single linear layer that takes the MLPF latent representations as input.
Any kinematic operations follow the natural output structure of each task and are held fixed.
For the jet-level tasks, the per-particle latent representations are averaged across the constituents of each jet before the linear layer produces the output.
For \pvecmiss regression, the linear layer produces a per-particle scalar weight, and the predicted \pvecmiss is the weighted vector sum of the PF candidate momenta in the event.
The \probe models have 129 parameters for the two regression tasks (128 weights $+$ 1 bias for the scalar output).
For jet flavor identification, which involves three output classes, the \probe model has 387 parameters ($128 \times 3$ weights $+$ 3 biases).
All \probe models are trained on the same data and with the same protocol as the other model variants.

While the \probe model quantifies the information that is linearly accessible in the latent space, we additionally study how much is accessible to a more expressive model by training \probeL, in which the full task-specific architecture is applied to the MLPF latent representations alone, with all other input features removed.
We observe that the \probeL model matches the \latents model for the two regression tasks and approaches the \baseline model for jet flavor identification, indicating that most of the task-relevant information is accessible from the latent space alone. Full results are reported in Appendix~\ref{app:nlp}.

The scaling figures (Figs.~\ref{fig:btagging:scaling}, \ref{fig:jer:scaling}, and \ref{fig:met:scaling}) are evaluated on the validation set and show the median (M) and interquartile range (IQR) across five random-seed trainings as a function of the size of the training set.
For jet flavor identification, we plot the AUC rather than the validation loss, as it is more directly interpretable.
The final physics performance results (Figs.~\ref{fig:btagging:roc}, \ref{fig:jer:physics}, and \ref{fig:met:physics}) use the held-out test set and the best model for each task (corresponding to the seed with the lowest validation loss when trained on the full training set), with shaded bands showing the statistical uncertainty per bin where applicable.


\section{Jet flavor identification}
\label{sec:btag}

We formulate jet flavor identification as a multi-class classification task where, given the PF candidates clustered into a jet, the model predicts one of three flavor categories: \PQb jets (from beauty quarks), \PQc jets (from charm quarks), or light-flavor jets (from strange, up, down quarks, or gluons).
The three-class setup separates the two heavy-flavor categories whose discrimination is the main target of modern flavor taggers.
We use the \ttbar sample described in Sec.~\ref{sec:setup:dataset}, which naturally provides all three classes: \PQb jets from top quark decays, \PQc jets from $\PW \to \PQc\overline{\PQs}$ decays, and light-flavor jets from hadronic \PW boson decays and final-state radiation.

\subsection{Jet clustering and labeling}
Reconstructed jets are clustered from the PF candidates using the generalized-$\kt$ algorithm~\cite{Cacciari:2008gp} with $R = 0.4$ and exponent $p = -1$, requiring a minimum $\pt > 5\GeV$, consistent with the convention used in Ref.~\cite{Mokhtar:2025mlpf}.
Jet-flavor labels are assigned directly to the reconstructed jets using the ghost-association technique~\cite{Cacciari:2008zb}, where generator-level \PB and \PD hadrons are added to the input list of the jet-clustering algorithm with infinitesimal momentum, so their presence assigns them to a jet without affecting the jet kinematics.
A reconstructed jet is labeled as a \PQb jet if any \PB hadron is associated with it in this way, a \PQc jet if no \PB hadron but at least one \PD hadron is associated, and a light-flavor jet otherwise.

\subsection{Dataset statistics}
A kinematic selection of $\pt \in [10, 200]\GeV$ and $|\eta| < 2.5$ is applied, retaining 89.3\% of jets.
Selected jets have a mean constituent multiplicity of 13 PF candidates and a \pt distribution with mean 42.2\GeV, median 39.4\GeV, and 90th percentile at 71.7\GeV.
The resulting flavor composition of the training set is approximately $51\%$ light-flavor jets, $13\%$ \PQc jets, and $36\%$ \PQb jets.
Table~\ref{tab:btag_splits} summarizes the jet dataset after the kinematic selection.
          
\begin{table}[htbp]
\centering
\begin{tabular}{l@{\hspace{1.em}}r@{\hspace{1.em}}r@{\hspace{1.em}}r@{\hspace{1.em}}r}
\toprule
Split      & Total jets     & Light-flavor jets & \PQc jets  & \PQb jets  \\
\midrule
Train      &  1{,}819{,}536 &   937{,}080       &  227{,}422 & 655{,}034  \\
Validation &    25{,}900    &   13{,}362        &  3{,}165   & 9{,}373  \\
Test       &   207{,}891    &  107{,}170        &  25{,}874  & 74{,}847  \\
\bottomrule
\end{tabular}
\caption{Jet counts for the jet flavor identification downstream task.}
\label{tab:btag_splits}
\end{table}

\subsection{Input features}
Each PF candidate is described by a set of kinematic, PID, and track-displacement variables, listed in Table~\ref{tab:btag_features}.
The kinematic features include the angular separations $\Delta\eta$ and $\Delta\phi$ with respect to the jet axis, $\log(1+\pt)$, $\log(1+E)$, $\log(\pt/\pt{_\mathrm{jet}})$, $\log(E/E_\mathrm{jet})$, and $\Delta R$, the angular distance to the jet axis.
Particle identification is encoded as a one-hot vector over the five MLPF output classes, together with the electric charge $q$.

\PB and \PD hadrons produced in heavy-flavor decays travel a macroscopic distance before decaying, leaving a measurable displacement between the primary vertex (PV) and the hadron decay point.
Exploiting this displacement is the cornerstone of heavy-flavor jet identification, and providing the model with track-level displacement information is standard practice for ML-based taggers.
In the MLPF reconstruction, each PF candidate is aligned with the input detector element from which it is reconstructed: a track for charged particles and a calorimeter cluster for neutral particles~\cite{Mokhtar:2025mlpf}.
For charged PF candidates, we therefore derive the full set of displacement variables listed in Table~\ref{tab:btag_features} from the associated track, following Ref.~\cite{Selvaggi:2022btag} and the numerical implementation of the official CLD flavor-tagging code~\cite{aumiller_2024_4pcr6-r0d06}: the transverse and longitudinal IPs ($d_0$, $z_0$), the signed 2D and 3D IPs and their significances ($\mathrm{SIP_{2D}}$, $\mathrm{SIP_{2D}}/\sigma_{2D}$, $\mathrm{SIP_{3D}}$, $\mathrm{SIP_{3D}}/\sigma_{3D}$), the 3D distance between the track and the jet axis at their point of closest approach ($d_\mathrm{3D}$) and its significance ($d_\mathrm{3D}/\sigma_{d_{3D}}$), and the full 15-element track parameter covariance matrix $C_{ij}$.
Neutral PF candidates have no associated track, so all displacement and covariance features are set to zero for them.

\begin{table}[htbp]
\centering
\begin{tabular}{l@{\hspace{0.5em}}l@{\hspace{0.5em}}l}
\toprule
Index & Feature & Description \\
\midrule
0--1   & $\Delta\eta$, $\Delta\phi$       & $\eta$, $\phi$ relative to jet axis \\
2      & $\log(1+\pt)$                    & Log-scaled \pt \\
3      & $\log(1+E)$                      & Log-scaled energy \\
4      & $\log(\pt/\pt{_\mathrm{jet}})$   & Log-scaled fraction of jet \pt \\
5      & $\log(E/E_\mathrm{jet})$         & Log-scaled fraction of jet energy \\
6      & $\Delta R$                       & Angular distance to jet axis \\
7      & $d_0$                            & Transverse IP \\
8      & $z_0$                            & Longitudinal IP \\
9      & $\mathrm{SIP_{2D}}$              & Signed 2D IP \\
10     & $\mathrm{SIP_{2D}}/\sigma_{2D}$  & Signed 2D IP significance \\
11     & $\mathrm{SIP_{3D}}$              & Signed 3D IP \\
12     & $\mathrm{SIP_{3D}}/\sigma_{3D}$  & Signed 3D IP significance \\
13     & $d_\mathrm{3D}$                  & Jet-track 3D distance \\
14     & $d_\mathrm{3D}/\sigma_{d_{3D}}$  & Jet-track 3D distance significance \\
15--29 & $C_{ij}$                         & Track parameter covariance matrix \\
30     & $q$                              & Electric charge \\
31--35 & PID flags                        & Five-class type encoding \\
\midrule
\multicolumn{3}{l}{\textit{Used by \latents and \probe only:}} \\
36--163 & Latent & 128-dim latent vectors \\
\bottomrule
\end{tabular}
\caption{
Per-particle input features for the jet flavor identification and jet energy regression tasks, following Ref.~\cite{Selvaggi:2022btag}.
}
\label{tab:btag_features}
\end{table}

Since the displacement variables require knowledge of the PV position, which is not provided by default in the CLD reconstruction chain, we use the \pythia generator-level PV smeared with a Gaussian distribution of width $3\unit{\mu m}$ in each spatial coordinate.
This corresponds to the expected vertex resolution of the FCC-ee collider design~\cite{Bacchetta:2021ibt} and serves as a stand-in for a full PV reconstruction, which we leave to future CLD reconstruction work.
The complete feature list is given in Table~\ref{tab:btag_features}.

\subsection{Model architecture and training}
We use ParticleNet~\cite{Qu:2019gqs} as a well-established architecture baseline, which allows the contribution of the MLPF latent representations to be isolated from model-specific gains.
We follow Ref.~\cite{Qu:2019gqs} for the architecture details, with EdgeConv channels 64/128/256, $\Delta\eta$ and $\Delta\phi$ as graph coordinates for the $k$-nearest-neighbor construction with $k = 16$, and a dropout rate of 0.1 in the post-pooling fully-connected head.
The parameter counts for the three model variants are summarized in Table~\ref{tab:btag_models}.
   
\begin{table}[htbp]
\centering
\begin{tabular}{l@{\hspace{1.2em}}l@{\hspace{1.2em}}c@{\hspace{1.2em}}c}
\toprule
Model & Architecture & Inputs & Parameters \\
\midrule
\baseline & ParticleNet & 36  & 360{,}651 \\
\latents  & ParticleNet & 164 & 508{,}363 \\
\probe    & Linear      & 128 & 387 \\
\bottomrule
\end{tabular}
\caption{
Model variants for the jet flavor identification task.
The \probe model is a single linear layer with 387 parameters ($128 \times 3$ weights + 3 biases), where the factor of 3 corresponds to the three flavor classes (\PQb, \PQc, light).
}
\label{tab:btag_models}
\end{table}

Models are trained on a heterogeneous pool of GPUs (NVIDIA A10, L40S, V100, RTX~3090, and RTX~4090) on the National Research Platform (NRP) Kubernetes cluster~\cite{Smarr:2018prp}, with wall-clock training time ranging from $\sim$14 minutes (smallest dataset, fastest GPU) to $\sim$9 hours (full dataset, on slower GPUs).
We use the AdamW optimizer~\cite{adamw} with an initial learning rate of $10^{-3}$, cosine annealing to $10^{-5}$, batch size 256, and a weight decay of $10^{-4}$.
We minimize the categorical cross-entropy loss with per-class weights set to the inverse training-set frequency to compensate for the imbalance between the three flavor classes.
The corresponding weights are approximately $(0.46, 1.88, 0.66)$ for light, \PQc, and \PQb jets, respectively.
Training is capped at 100 epochs with early stopping after 20 epochs without improvement in validation loss.

To characterize the data-scaling behavior, we train on five training-set sizes: 100k, 200k, 500k, 1M, and the full 1.82M jets.


\subsection{Results}
\label{sec:tasks:btag:results}

\begin{figure}[htbp]
  \centering                                            
  \includegraphics[width=0.42\textwidth]{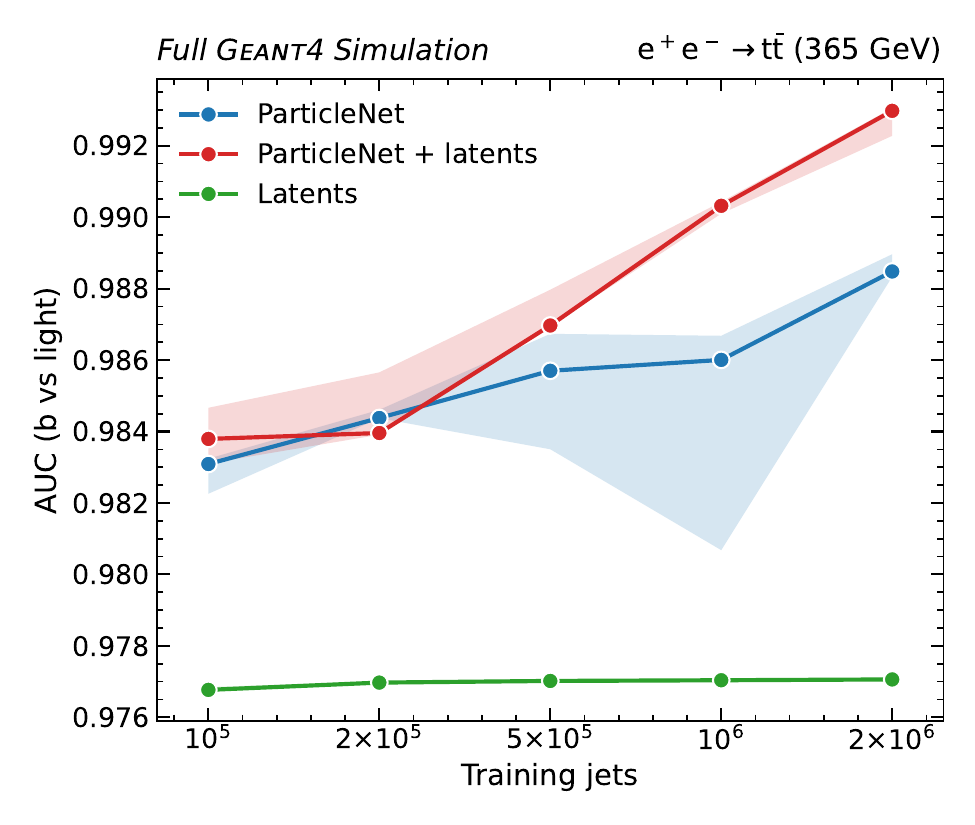}
  \includegraphics[width=0.42\textwidth]{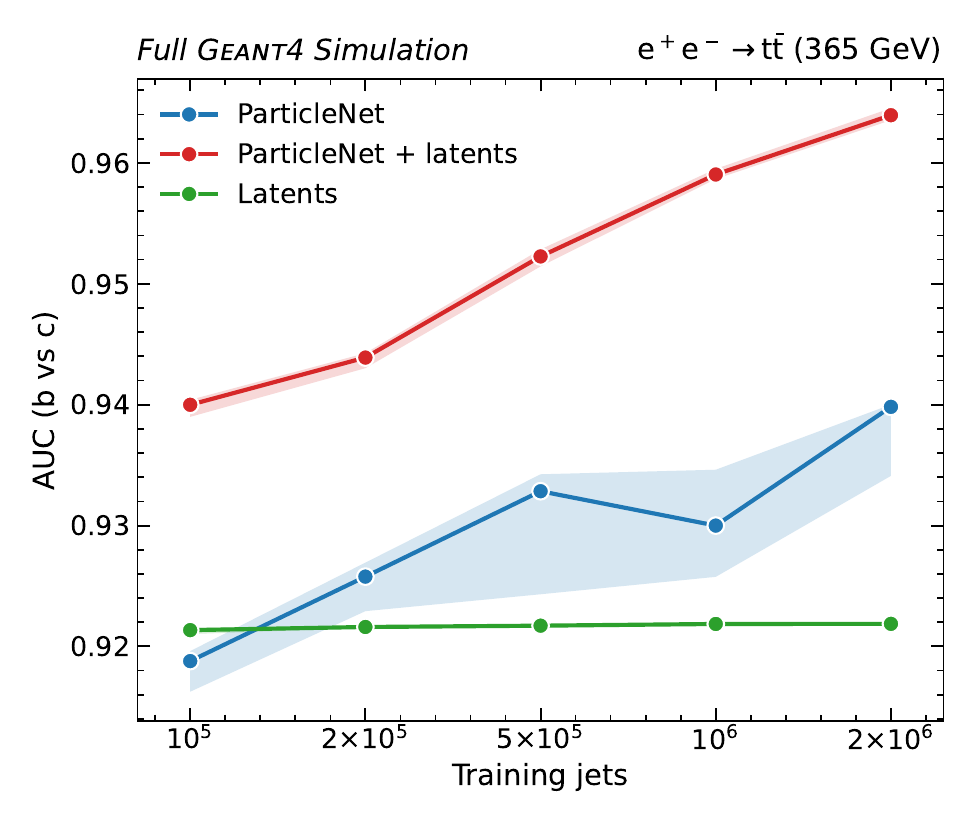}
  \caption{
    AUC for \PQb-vs-light (top) and \PQb-vs-\PQc (bottom) discrimination as a function of the number of training jets, comparing the three model variants (blue: \baseline; red: \latents; green: \probe), evaluated on the validation set over all reconstructed jets ($p_\mathrm{T}>10$ \GeV, $|\eta|<2.5$).
    Each curve shows the median across five random seeds, and the shaded bands indicate the interquartile range.
  }
  \label{fig:btagging:scaling}
\end{figure}

To evaluate the classification performance, we use the receiver operating characteristic (ROC) curve, which reports the jet mis-identification rate as a function of \PQb-jet identification efficiency, and the corresponding area under the curve (AUC).
Figure~\ref{fig:btagging:scaling} shows the AUC at the best validation epoch, evaluated on the held-out validation set, for \PQb-vs-light (top) and \PQb-vs-\PQc (bottom) discrimination, as a function of the number of training jets.

The \latents model dominates both AUC scaling curves at every training-set size, with the advantage more pronounced for the harder \PQb-vs-\PQc discrimination.
Trained on only 100k jets, it already reaches an AUC of approximately 0.940 for \PQb-vs-\PQc, almost the same value as the AUC of the \baseline model trained on the full training set.
The \baseline model is unable to close the AUC gap for both \PQb-vs-light and \PQb-vs-\PQc discrimination at the largest available training size, indicating that the MLPF latent representations supply information that is not present in the baseline input feature set, despite the latter already including track-level discriminant variables.

The \probe curve is essentially flat on both plots across the entire range, saturating at AUC $\approx 0.977$ for \PQb-vs-light and $\approx 0.922$ for \PQb-vs-\PQc and showing no improvement with training-set size.
This flatness is expected by construction.
A linear head has no nonlinear capacity to fit, so its performance reflects only the discrimination power already encoded in the frozen latent representations.
However, the \probe model achieves a nontrivial AUC well above random chance despite the fact that the MLPF backbone never uses jet-flavor labels during training, indicating that flavor-discriminating structure is intrinsically present in the latent representations.
At the full training-set size the \probe model trails the \baseline model by approximately $0.012$ in AUC for \PQb-vs-light and $0.018$ for \PQb-vs-\PQc.

\begin{figure}[htbp]
  \centering
  \includegraphics[width=0.47\textwidth]{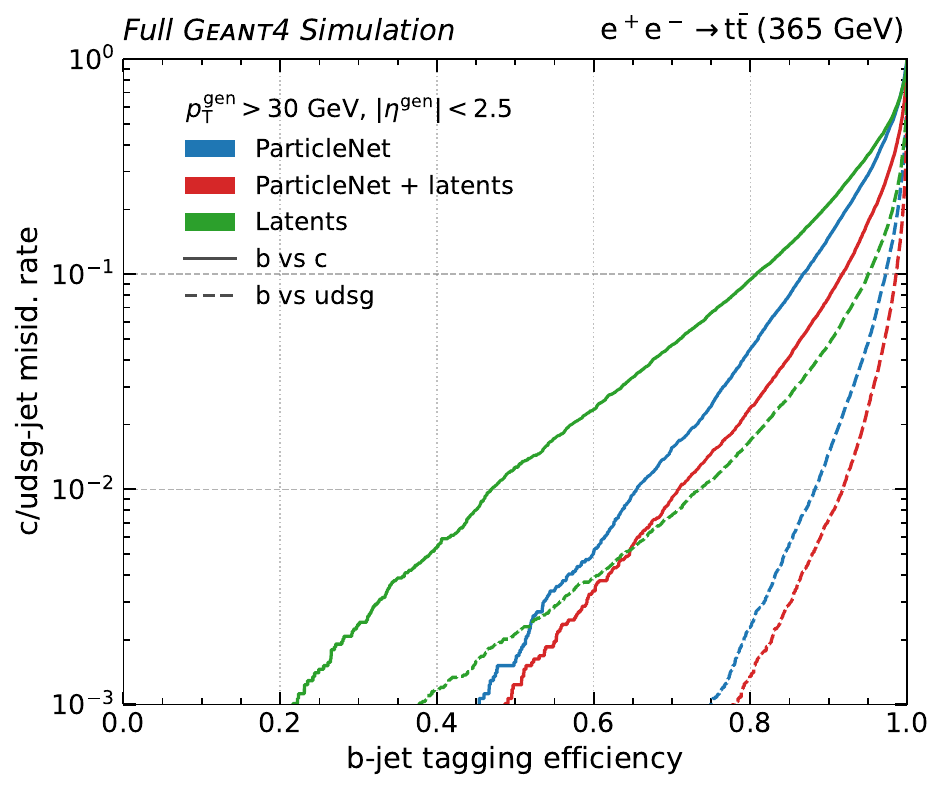}
  \caption{
        ROC curves for the three model variants trained on the full training dataset and evaluated on the test set using the best model (the lowest-validation-loss seed) for each variant.
        Shown are the mis-identification rates for light-flavor jets (dashed) and \PQc jets (solid) as a function of \PQb-jet identification efficiency, with a fiducial selection of $\pt^{\mathrm{gen}} > 30\GeV$ and $|\eta^{\mathrm{gen}}| < 2.5$.
  }
  \label{fig:btagging:roc}
\end{figure}

In Fig.~\ref{fig:btagging:roc}, we present the ROC curves for the three model variants trained on the full training dataset and evaluated on the test set with a fiducial kinematic selection of $\pt^{\mathrm{gen}} > 30\GeV$ and $|\eta^{\mathrm{gen}}| < 2.5$, separately for the \PQb-vs-light (dashed) and \PQb-vs-\PQc (solid) cases.

At a typical working point of 1\% mis-identification rate, the \latents model improves the \PQb-jet identification efficiency by approximately 3\% against light-flavor jets and approximately 6\% against \PQc jets, relative to the \baseline model.
At the same working point, the \probe model shows \PQb-jet identification efficiency losses of approximately $15\%$ against light-flavor jets and $19\%$ against \PQc jets, relative to the \baseline model.
This gap reflects the substantial discriminating power supplied by the explicit track-displacement features in the \baseline model input set, which a linear projection of the latent space alone cannot fully recover.

\section{Jet energy regression}
\label{sec:jer}

We formulate jet energy regression as a per-jet scalar regression task where, given the PF candidates clustered into a jet, the model predicts a multiplicative correction factor that maps the reconstructed jet energy to the corresponding generator-level jet energy.
This is the form adopted in recent ML-based jet energy regression studies at CMS using ParticleNet-based architectures~\cite{CMS-DP-2024-066}.

We use the same jet dataset (Table~\ref{tab:btag_splits}), and input features (Table~\ref{tab:btag_features}) as the jet flavor identification task, including the track-displacement variables.
Although these variables are not required for jet energy regression, retaining them makes the model flavor-aware and follows the convention adopted in Ref.~\cite{CMS-DP-2024-066}.
The task-specific differences from jet flavor identification are the regression target, the loss function, and the output head, described in the following subsections.

\subsection{Regression target and loss}
For each reconstructed jet we form the multiplicative correction factor,
\begin{equation}                                  
    y = \frac{E^{\mathrm{gen}}_{\mathrm{jet}}}{E^{\mathrm{reco}}_{\mathrm{jet}}},
\end{equation}
where $E^{\mathrm{reco}}_{\mathrm{jet}}$ is the energy of the reconstructed jet and $E^{\mathrm{gen}}_{\mathrm{jet}}$ is the energy of its matched generator-level jet.
Generator-level jets are obtained by applying the same generalized-$k_t$ algorithm to stable particles at generator level, and are matched to reconstructed jets within $\Delta R < 0.4$.
Reconstructed jets without a generator-level match within $\Delta R < 0.4$ are excluded from the sample.

We use the log-cosh loss,
which is approximately quadratic for small residuals and approximately linear for large ones.
This loss profile is robust to the heavy tail of rare mis-reconstructed jets without sacrificing sensitivity to small residuals.
Predicting the ratio rather than the generator-level energy decouples the target from the absolute \pt scale, which prevents the loss from being dominated by the large absolute residuals of high-\pt jets and keeps the log-cosh objective in its quadratic regime across the full \pt range.

\subsection{Model architecture and training}
The architecture and training protocol are identical to those of the jet flavor identification task (Sec.~\ref{sec:btag}), with the only change being that the final
head outputs a scalar prediction instead of the three-class output.
For the \probe model, we initialize the linear head with zero weights and unit bias so the untrained model outputs the constant value $1.0$ for every jet, close to the empirical target mean of $1.02$.
Training then measures the additional information that can be extracted by a linear readout of the MLPF latent representations.

The parameter counts are $360{,}137$ for the \baseline model, $507{,}849$ for the \latents model, and $129$ for the \probe model, slightly smaller than the corresponding jet flavor identification counts (Table~\ref{tab:btag_models}) because the output head produces a single scalar prediction rather than three flavor outputs.

To characterize the data-scaling behavior, we train on the same five training-set sizes as jet flavor identification: 100k, 200k, 500k, 1M, and the full 1.82M jets.

\subsection{Results}
\begin{figure}[htbp]
  \centering
  \includegraphics[width=0.39\textwidth]{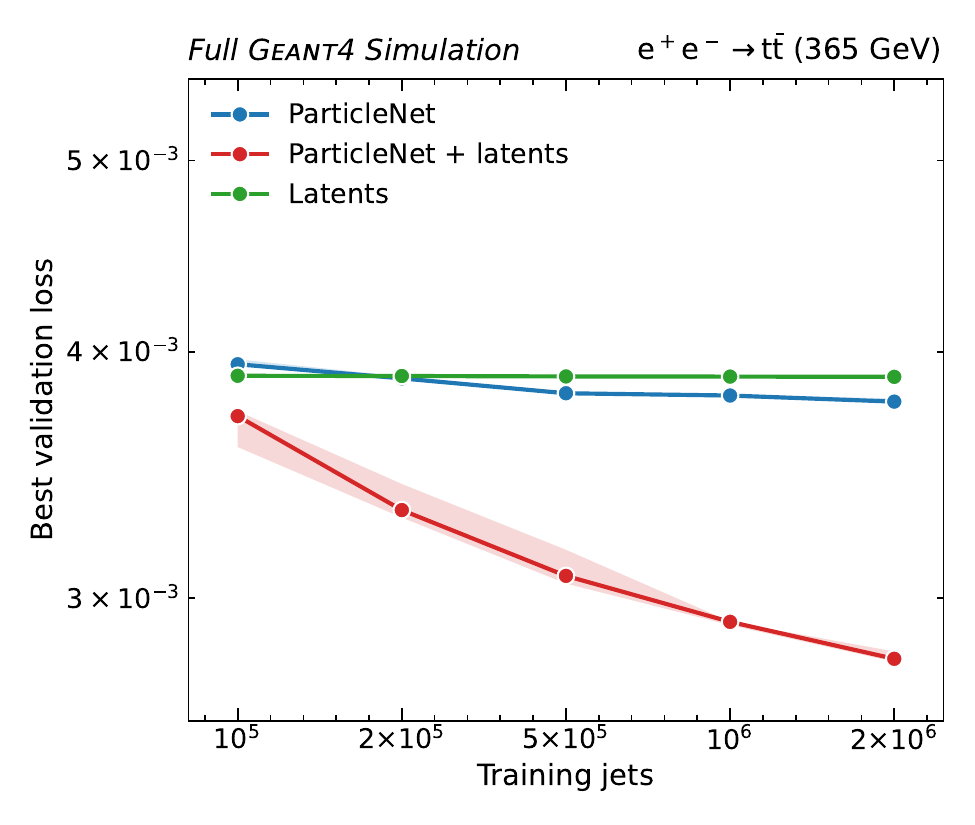}
  \caption{
      Best validation loss as a function of the number of training jets, for the jet energy regression task, comparing the three model variants (blue: \baseline; red: \latents; green: \probe), evaluated on the validation set.
      Each curve shows the median across five random seeds, and the shaded bands indicate the interquartile range.
  }
  \label{fig:jer:scaling}
\end{figure}

We first study how performance scales with the size of the training set.
Figure~\ref{fig:jer:scaling} shows the best validation loss as a function of the number of training jets, evaluated on the held-out validation set.
The data-scaling behavior is similar to that observed in the jet flavor identification task.
The \latents model dominates at every training-set size.
As in the jet flavor identification case, the \probe curve is essentially flat by construction throughout the full range, plateauing slightly above the \baseline curve at larger training sizes.

\begin{figure*}[htbp]
  \centering
  \includegraphics[width=0.46\textwidth]{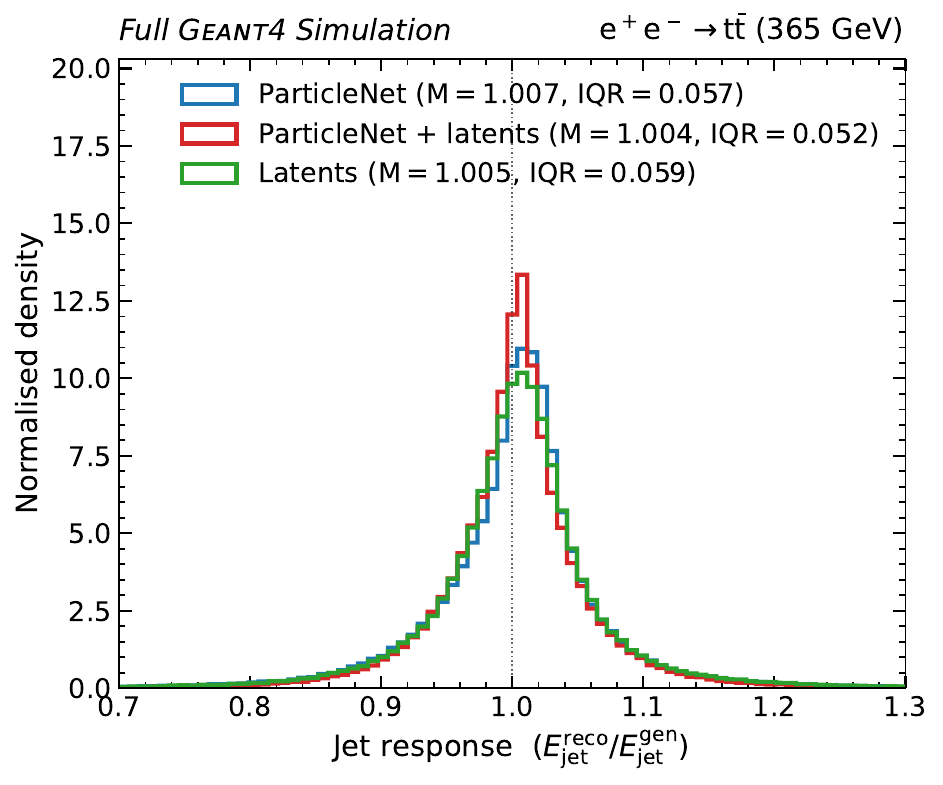}
  \includegraphics[width=0.46\textwidth]{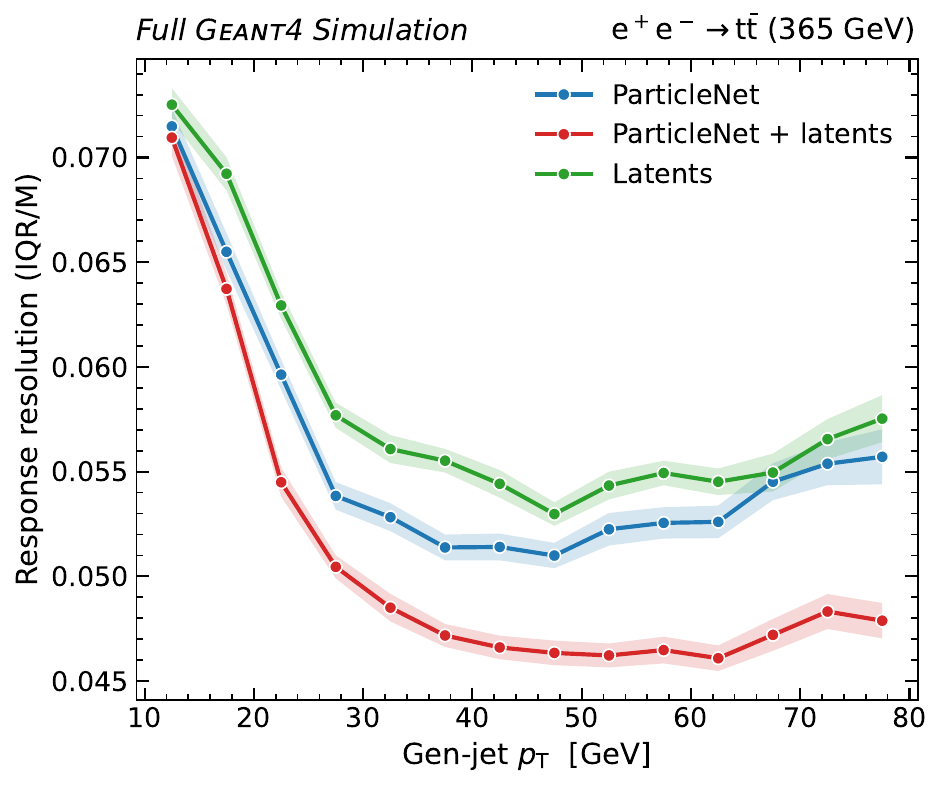}
  \caption{
      Jet energy performance at the largest training-set size (blue: \baseline; red: \latents; green: \probe), evaluated on the test set using the best model (the lowest-validation-loss seed) for each variant.
      Left: inclusive distribution of the jet energy response.
      Right: jet energy resolution as a function of the generator-level jet \pt.
      Shaded bands indicate the per-bin statistical uncertainty.
  }
  \label{fig:jer:physics}
\end{figure*}

We evaluate the physics performance using the jet energy response, defined as $E^{\mathrm{reco}}_{\mathrm{jet}} / E^{\mathrm{gen}}_{\mathrm{jet}}$, where the corrected reconstructed-jet energy is obtained by multiplying the raw reconstructed energy by the model's predicted correction factor.
The inclusive response distribution at the full training set is shown in the left panel of Fig.~\ref{fig:jer:physics}.
All three distributions are centered close to unity, with the \latents model achieving the tightest distribution with an IQR of $0.052$ against $0.057$ for the \baseline model, an improvement of approximately 10\%.
The \probe distribution sits at an IQR of $0.059$, trailing the \baseline by approximately 3\%.

The right panel of Fig.~\ref{fig:jer:physics} shows the jet energy resolution, defined as IQR/M, as a function of the generator-level jet \pt.
The resolution improves rapidly with increasing \pt for all three models, as expected.
The \latents model achieves the best resolution at every \pt bin, with a relative improvement over the \baseline model of approximately 10\% at intermediate \pt and approximately 15\% at the highest \pt bins.
The \probe model trails the \baseline model by approximately 5--8\% at intermediate \pt but approaches it at the highest \pt bins.

\section{Missing momentum regression}
\label{sec:met}

The third downstream task we consider is the regression of \pvecmiss , which represents the imbalance of the visible transverse and longitudinal momenta and is a key observable for physics analyses involving escaping neutral particles such as neutrinos.
At the LHC, ML-based missing transverse momentum (\ptmiss) estimators have become standard tools in precision analyses.
For instance, the recent CMS \PW boson mass measurement~\cite{CMS:Wmass} relied on the DeepMET algorithm~\cite{CMS:DeepMET} to estimate the \ptmiss, which improves the reconstruction of the \PW boson.
The FCC-ee environment is pileup-free, but the same fundamental challenge persists.
The \ptmiss must be inferred from a list of reconstructed PF candidates whose individual momentum estimates carry detector-resolution uncertainties that accumulate when summed over the event.

Unlike at hadron colliders, where the longitudinal boost of the partonic initial state is unknown and only the \ptmiss is well-defined, the $\PGe^+\PGe^-$ initial state has fixed, known kinematics.
The missing momentum vector \pvecmiss is therefore physically meaningful and gives direct access to observables such as the missing mass, which is central to several FCC-ee physics analyses~\cite{FCC-ee}.
We accordingly extend the DeepMET regression target from the transverse plane to the full three-dimensional missing momentum.
The remainder of this section defines the input features, the DeepMET-based \baseline architecture, the training procedure, and the resulting physics performance.

\subsection{Dataset statistics}
We use the inclusive \ttbar sample described in Sec.~\ref{sec:setup:dataset}, which naturally provides a mix of events with genuine missing momentum from neutrinos in leptonic \PW boson decays and events with no true missing momentum from fully hadronic final states.
We make no distinction between these classes in the performance plots.
The \ptmiss distribution over the validation set has a mean of 31.1\GeV, a median of 23.4\GeV, and a 95th percentile at 82.7\GeV.
We therefore restrict the range of the physics performance plots to 100\GeV.

\subsection{Input features}
Each PF candidate is described by eight continuous features and two categorical features.
The continuous features are \pt, $E$, $\eta$, the Cartesian momentum components $p_x$, $p_y$ and $p_z$, and the IPs ($d_0$ and $z_0$) relative to the reconstructed PV (Sec.~\ref{sec:btag}).     
The two kinematic variables \pt and $E$ enter through a $\log(1+x)$ transformation to soften their heavy tails; $p_x$, $p_y$ and $p_z$ retain their physical units since they enter the \pvecmiss kinematic sum directly.
The categorical features are the PID class (one of the five MLPF output classes) and the electric charge, amounting to 14 input features per PF candidate.

\subsection{Model architecture}
We implement the DeepMET architecture in \textsc{PyTorch}~\cite{pytorch}, closely following the original CMS implementation~\cite{CMS:DeepMET}.
The network is a per-particle multilayer perceptron applied independently to each PF candidate, with hidden dimensions $[64, 32, 16]$ and batch normalization and ReLU activations after each fully-connected layer.
For each PF candidate the network outputs four quantities: a scalar weight $w_i$ shared across the three spatial axes, and three bias terms $b_{i,x}$, $b_{i,y}$, and $b_{i,z}$.
The event-level prediction is then assembled as a masked weighted sum along each axis,
\begin{align}
    \hat{p}^{\mathrm{miss}}_x &= - \sum_{i} \left( w_i \, p_{i,x} + b_{i,x} \right), \\
    \hat{p}^{\mathrm{miss}}_y &= - \sum_{i} \left( w_i \, p_{i,y} + b_{i,y} \right), \\
    \hat{p}^{\mathrm{miss}}_z &= - \sum_{i} \left( w_i \, p_{i,z} + b_{i,z} \right),
\end{align}
where the sum runs over the PF candidates in the event.
We adapt the original CMS DeepMET architecture to the $\PGe^+\PGe^-$ environment in two ways.
We omit the pileup-per-particle identification (PUPPI)~\cite{Bertolini:2014bba} weight feature, since the $\PGe^+\PGe^-$ environment is pileup-free.
We replace the PF candidate mass input with the PF candidate energy, since the mass of the PF candidate is set by its identified particle species and is therefore redundant with the PID embedding, whereas the energy carries independent kinematic information.
The parameter counts for the three model variants are summarized in Table~\ref{tab:met_models}.

\begin{table}[htbp]
\centering
\begin{tabular}{l@{\hspace{1.2em}}l@{\hspace{1.2em}}c@{\hspace{1.2em}}c}
\toprule
Model & Architecture & Inputs & Parameters \\
\midrule
\baseline & DeepMET  & 14   & 4{,}564 \\
\latents  & DeepMET  & 142 & 12{,}756 \\
\probe    & Linear   & 128 & 129 \\
\bottomrule
\end{tabular}
\caption{
Model variants for the missing momentum regression task.
The \probe model is a single linear layer with 129 parameters (128 weights + 1 bias).
}
\label{tab:met_models}
\end{table}

\subsection{Training}
We train the model to minimize the mean squared error between the predicted and generator-level \pvecmiss vectors,
\begin{multline}
  \mathcal{L} = \frac{1}{2}\,\Big\langle\,
    (\hat{p}^{\mathrm{miss}}_x - p^{\mathrm{miss,gen}}_x)^2
  + (\hat{p}^{\mathrm{miss}}_y - p^{\mathrm{miss,gen}}_y)^2 \\
  + (\hat{p}^{\mathrm{miss}}_z - p^{\mathrm{miss,gen}}_z)^2
  \,\Big\rangle.
\end{multline}

Models are trained on the same heterogeneous GPU pool as the other downstream tasks, with wall-clock training time ranging from $\sim$30 minutes (smallest dataset, fastest GPU) to $\sim$13 hours (full dataset, slower GPUs).
We use the AdamW optimizer~\cite{adamw} with a constant learning rate of $10^{-3}$, no weight decay, and batch size 128.
Training is capped at 100 epochs with early stopping after 20 epochs without improvement in validation loss.

To characterize the data-scaling behavior, we train on five training-set sizes: 10k, 50k, 100k, 200k, and the full 350k events.

\subsection{Results}
\begin{figure}[htbp]
  \centering
  \includegraphics[width=0.36\textwidth]{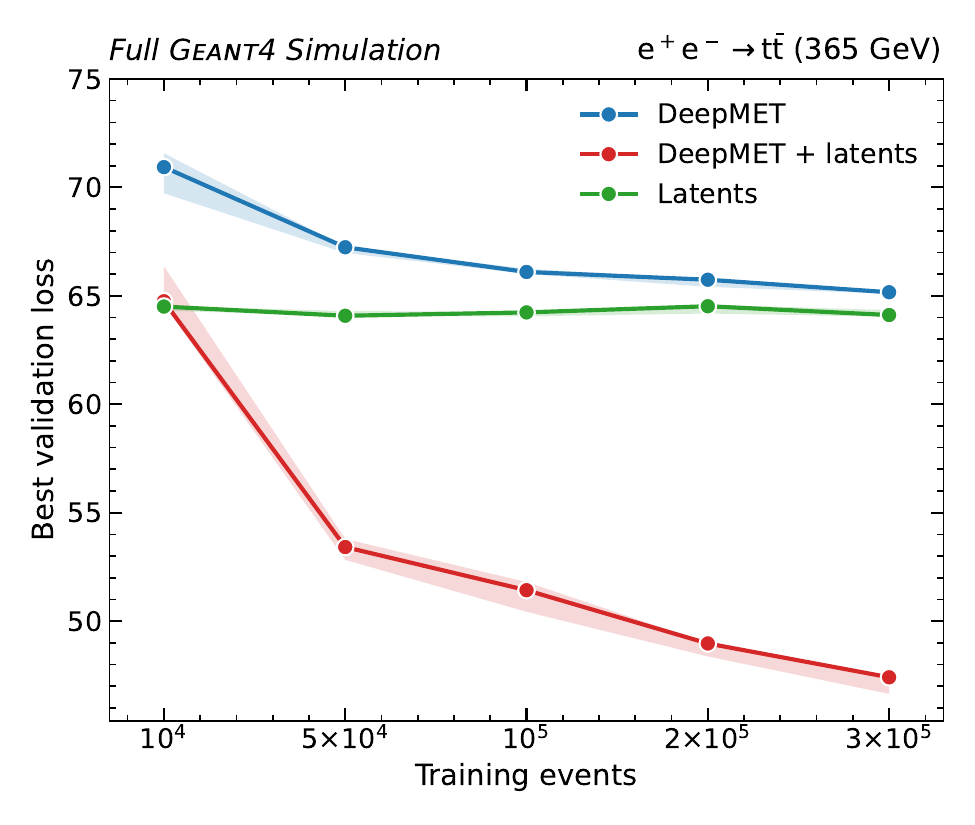}
  \caption{        
      Best validation loss as a function of the training dataset, for the missing momentum regression task, comparing the three model variants (blue: \baseline; red: \latents; green: \probe), evaluated on the validation set.
      Each curve shows the median across five random seeds, and the shaded bands indicate the interquartile range.
  }
  \label{fig:met:scaling}
\end{figure}

Figure~\ref{fig:met:scaling} shows the best validation loss, evaluated on the held-out validation set, as a function of the number of training events.
The \latents model reaches a validation loss of approximately 47.9 at the full training set, compared with approximately 65.0 for the \baseline model, a 26\% improvement.
The \latents model trained on only 50k events already beats the \baseline model trained on the full sample, indicating that the MLPF latent representations supply information of central importance to the \pvecmiss regression task.
The \probe curve is essentially flat across the full range of training-set sizes, plateauing near 64.2.

\begin{figure*}[htbp]
\centering
\includegraphics[width=0.32\textwidth]{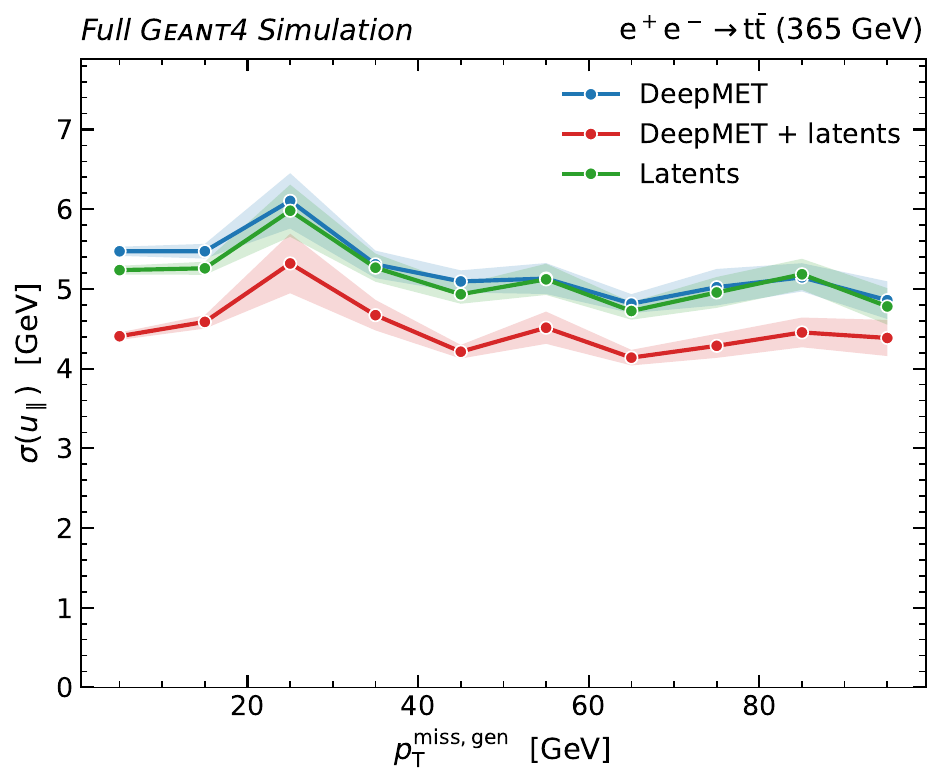}
\hfill
\includegraphics[width=0.32\textwidth]{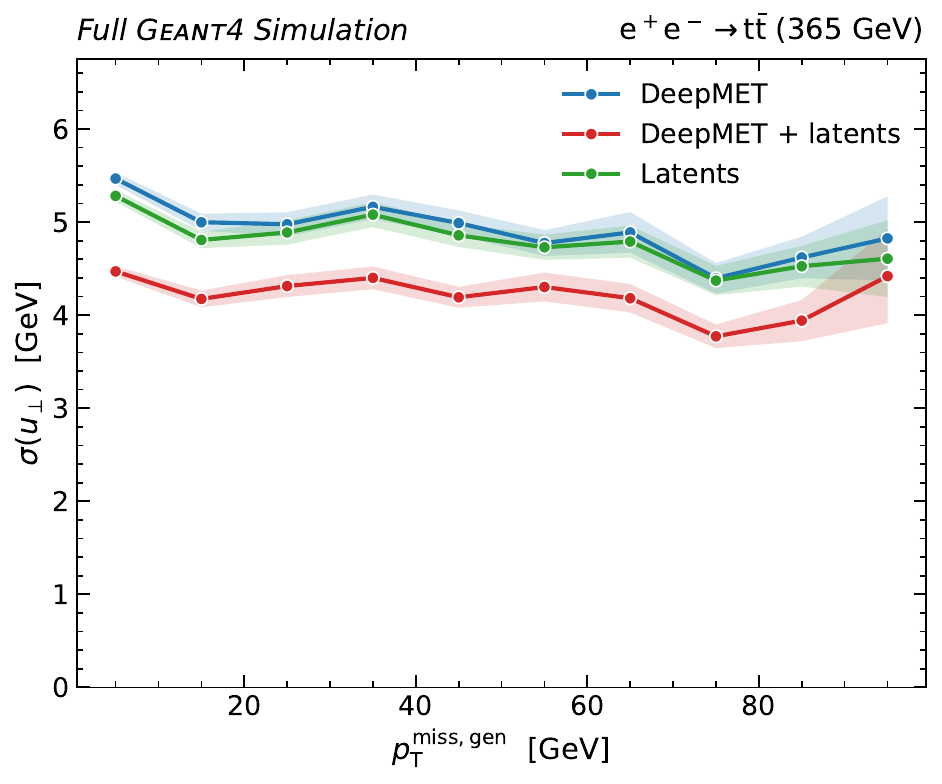}
\hfill
\includegraphics[width=0.32\textwidth]{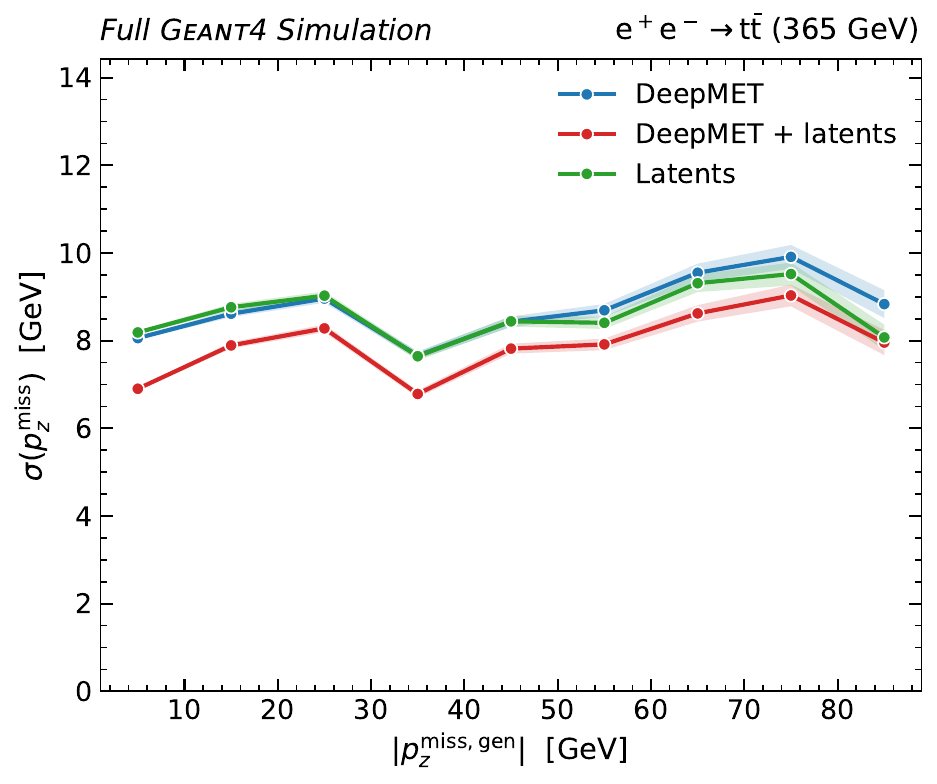}
\caption{
    Missing momentum performance at the largest training-set size (blue: \baseline; red: \latents; green: \probe), evaluated on the test set using the best model (the lowest-validation-loss seed) for each variant.
    Left, middle: recoil resolution parallel ($\sigma(u_\parallel)$) and perpendicular ($\sigma(u_\perp)$) to the generator-level \ptmiss direction, as a function of the \ptmiss magnitude.
    Right: longitudinal missing-momentum resolution $\sigma(p_z^{\mathrm{miss}})$ as a function of $|p_z^{\mathrm{miss,gen}}|$.
    Shaded bands indicate the per-bin statistical uncertainty.
}
\label{fig:met:physics}
\end{figure*}

For the physics performance we first inspect the inclusive distribution of the \ptmiss response, defined as the ratio of the reconstructed \ptmiss to its generator-level value $\ptmissReco / \ptmissGen$.
All three distributions are centered slightly below unity, with medians of $0.967$, $0.974$, and $0.962$ for the \baseline, \latents, and \probe models, respectively.

Following the methodology established in Ref.~\cite{CMS:DeepMET}, we further evaluate the recoil resolution in two orthogonal projections.
The recoil $\vec{u} = -\vec{p}^{\,\mathrm{miss,reco}}_\mathrm{T}$ is decomposed into components parallel ($u_\parallel$) and perpendicular ($u_\perp$) to the generator-level \ptmiss direction.
The parallel resolution $\sigma(u_\parallel)$ measures how well the recoil tracks the generator-level \ptmiss in magnitude, while the perpendicular resolution $\sigma(u_\perp)$ measures how well the direction is reconstructed.
These are reported in bins of the generator-level \ptmiss magnitude in the left and middle panels of Fig.~\ref{fig:met:physics}.
The \latents model improves the recoil resolution by approximately 15--20\% across the full range in both projections, while the \probe model gives a smaller but consistent improvement of approximately 2--5\% over the \baseline model.
The longitudinal resolution $\sigma(p_z^{\mathrm{miss}})$, defined as the standard deviation of the per-event residual $(p_z^{\mathrm{miss,gen}} - \hat{p}_z^{\mathrm{miss}})$ and shown in the right panel of Fig.~\ref{fig:met:physics}, follows the same hierarchy with the \latents model improving over the \baseline model by approximately 10\% across the full range.

\section{Discussion}
\label{sec:discussion}

The three downstream studies presented in Secs.~\ref{sec:btag}--\ref{sec:met} together support the central claim of this work.
The three tasks are qualitatively different: jet flavor identification, a per-jet multi-class classification that exploits track-level displacement variables; jet energy regression, a per-jet scalar regression; and missing momentum regression, a per-event vector regression.
In all three cases, augmenting the task-specific baseline with the frozen MLPF latent representations yields a consistent improvement on the physics-performance metric appropriate to each task, despite the three tasks differing in structure, the input modalities they emphasize, and their loss objectives.
For jet flavor identification, at a $1\%$ jet mis-identification rate, the \latents model improves the \PQb-jet identification efficiency by approximately 3\% against light-flavor jets and 6\% against \PQc jets.
For jet energy regression, the \latents model improves the jet energy resolution by approximately 10--15\% across the full jet \pt range, with the largest gains at the highest jet \pt.
The greatest improvement is observed for missing momentum regression, where the recoil resolution improves by approximately 15--20\% in both projections, and the longitudinal resolution improves by approximately 10\%, across the full range.

The \probe model results strengthen the foundation-model interpretation by quantifying the fraction of task-relevant information that is already linearly accessible in the MLPF latent space, even though the MLPF backbone was never trained using jet-flavor labels, generator-level jet energies, or generator-level \pvecmiss information.
The hierarchy varies by task, and the pattern itself is a useful diagnostic.
For missing momentum regression, a 129-parameter linear layer outperforms the DeepMET-based \baseline model at every training-set size while using approximately 35 times fewer parameters.
For the two jet-level tasks, the \probe model trails the \baseline model by approximately $3\%$ in the inclusive jet-energy resolution and more substantially in jet flavor identification.
Two factors contribute to this gap.
First, the \baseline model input set includes explicit track-displacement features---the IP significances and the signed 2D and 3D IPs---that are highly discriminating for jet flavor identification but were not part of the MLPF input.
MLPF sees only the raw perigee impact parameters $d_0$ and $z_0$, so these engineered features are not encoded in the latent representations, and a linear projection of the latent space cannot recover their discriminating power.
Second, the \probe model uniformly averages the latent representations across jet constituents, diluting constituent-level structure that the ParticleNet-based \baseline can exploit through its learned aggregation.
A more task-specific aggregation of the latent representations could further close the gap.

The scaling of the performance as a function of dataset size provides further evidence that the performance gap reflects information content rather than model capacity or training efficiency.
For the jet energy regression and missing momentum estimation tasks, the \baseline model performance saturates at large training-set sizes and the gap with the \latents model does not close with additional labeled data.
The capacity-control study of Appendix~\ref{app:capacity} confirms that this saturation is not a consequence of model size, and the relevant information is therefore not recoverable from the task-specific input features regardless of training-set size or model capacity.
For jet flavor identification, both models continue to improve with dataset size, yet the gap between them persists.
This persistence suggests that the advantage of the MLPF latent representations reflects information content rather than data efficiency.
The backbone encodes flavor-relevant physics that remains inaccessible to the downstream model through particle-level features alone.

To understand the physics mechanisms driving the improvements from the latent representations, we perform several checks.
For jet flavor identification, we zero the latent representations of neutral PF candidates at inference time for the \latents model and find that both discriminations degrade, with the larger drop observed for \PQb-vs-\PQc.
With a fiducial selection of $\pt^{\mathrm{gen}} > 30\GeV$ and $|\eta^{\mathrm{gen}}| < 2.5$, the AUC evaluated on the test set decreases from $0.994$ to $0.993$ for \PQb-vs-light and from $0.970$ to $0.961$ for \PQb-vs-\PQc, corresponding to a $\Delta\mathrm{AUC}$ of approximately $0.001$ and $0.009$, respectively.
This is consistent with the known physics of the two tasks.
For \PQb-vs-light, large-IP tracks in light-flavor jets arise from fake tracks, resolution effects, and some long-lived hadrons, so the explicit IP features in the \baseline model already provide strong discrimination.
For \PQb-vs-\PQc, both jet classes carry genuine secondary vertices (from particles with similar masses and lifetimes) and large-IP tracks, so in addition to IP-based discrimination the differences in quark fragmentation are also important.
The $\PQb\to\PQc\to\mathrm{daughters}$ cascade produces neutral activity across multiple decay steps, and we suspect that the global attention mechanism in MLPF encodes the collective neutral deposits from the full cascade jointly, a signal that is absent from the per-particle PF candidate interface and that the downstream model cannot reconstruct from kinematic features alone.

For missing momentum regression, we inspect the mean scalar weight that the \probe model assigns to each PID class over the test set. 
The hierarchy follows the expected ordering of per-particle momentum reconstruction reliability, where charged hadrons receive the highest mean weight ($\bar{w} = 0.995$), neutral hadrons the lowest ($\bar{w} = 0.902$), and electrons sit notably below muons ($\bar{w} = 0.926$ versus $0.973$), consistent with bremsstrahlung-induced reconstruction losses degrading per-particle momentum accuracy.
The weight standard deviation is highest for neutral hadrons ($\sigma_w = 0.117$) and lowest for muons ($\sigma_w = 0.042$), indicating that the \probe model varies weights at the per-particle level rather than applying a fixed per-class correction.
This hierarchy suggests that MLPF encodes per-particle reconstruction reliability in the latent representations.

For jet energy regression, we find that the improvement is uniform across jet flavors (approximately 9--11\% for all three classes) and shows no significant trend with constituent multiplicity, consistent with the latent representations supplying generic per-particle reconstruction-quality information rather than a topology-specific signal.

\section{Conclusion}
\label{sec:conclusion}

Combined with the cross-detector transfer-learning study of Ref.~\cite{Mokhtar:2025mlpf}, this work establishes a two-axis foundation-model picture for machine learning (ML)-based event reconstruction.
The machine-learned particle flow (MLPF) algorithm transfers across detector geometries with substantially less data than training from scratch, and the representations it learns transfer across qualitatively different downstream analysis tasks without retraining.
These two dimensions together motivate treating reconstruction models as natural foundation models for collider physics, useful both for detector R\&D and for downstream analysis.

The practical implication for analysis pipelines is direct.
Modern collider analyses devote significant effort to feature engineering and to training task-specific ML architectures, often on dataset slices that are themselves expensive to generate, simulate, or label.
In jet flavor identification, a 387-parameter linear classifier using only the MLPF latent representations achieves nontrivial discrimination without any hand-crafted track impact parameter features, suggesting that the reconstruction model already encodes flavor-relevant structure that would otherwise require careful feature engineering to exploit.

A similar implication holds for missing momentum regression.
For instance, the recent CMS \PW boson mass measurement~\cite{CMS:Wmass} relies on a DeepMET estimator, which was trained on a few tens of thousands of events with performance reported to be only weakly sensitive to larger training samples~\cite{CMS:DeepMET}.
An estimator built on MLPF latent representations could improve the performance at fixed data scale with no modification to the upstream reconstruction, and would continue to improve at larger training-set sizes where the \baseline model shows clear signs of saturation.
The \ptmiss study presented here is in a pileup-free environment, so the translation to LHC analyses with significant pileup remains to be tested, but the gains observed are large enough to motivate that test.

\section*{Acknowledgments}
F.M. and J.D. are supported by the Research Corporation for Science Advancement (RCSA) under Grant No. CS-CSA-2023-109, U.S. Department of Energy (DOE), Office of Science, Office of High Energy Physics under Grant No. DE-SC0009919, and the U.S. National Science Foundation (NSF) Harnessing the Data Revolution (HDR) Institute for Accelerating AI Algorithms for Data Driven Discovery (A3D3) under Cooperative Agreement No. PHY-2117997.
J.P. is supported by the Estonian Research Council grant PSG864, TARISTU24-TK10 and by the European Regional Development Fund through the CoE program grant TK202.
M.K. is supported by the U.S. Department of Energy (DOE) under Grant No. DE-AC02-76SF00515.
The authors contributed according to the contributor roles taxonomy (CRediT) categories as follows. 
F.M.: conceptualization, investigation, methodology, software, validation, data curation, writing (original draft), writing (review and editing).
J.P.: conceptualization, validation, supervision, writing (review and editing).
M.K.: conceptualization, supervision, writing (review and editing).
J.D.: conceptualization, resources, funding acquisition, supervision, writing (review and editing).
The authors acknowledge the use of Anthropic's Claude (Opus 4.7 and 4.8) for assistance with code development and manuscript editing.
The authors verified all AI-assisted output and take full responsibility for the scientific content of this paper.

\section*{Data availability}
The code and data that support the findings of this article are openly available~\cite{Mokhtar2026}.

\appendix

\section{Capacity control with \baselineL}
\label{app:capacity}

For each downstream task, the \latents model has more parameters than the \baseline model because it receives an additional 128-dimensional MLPF latent vector per particle as input.
To verify that the performance gain attributed to the latent representations is not simply a consequence of the larger parameter count, we train an additional model variant, \baselineL, with the same inputs as the \baseline model but with wider hidden layers chosen to match the parameter count of the \latents model.
The \baselineL model is trained with a single random seed per training-set size as a capacity check, and the comparison below uses the best model (i.e., the one trained on the full training dataset).

For jet flavor identification, \baselineL widens the ParticleNet EdgeConv channels from $(64, 128, 256)$ to $(80, 160, 320)$, raising the parameter count from $360{,}651$ to $510{,}955$, approximately matching the $508{,}363$ parameters of the \latents model.
For jet energy regression, \baselineL widens the same EdgeConv channels, raising the parameter count from $360{,}137$ to $510{,}441$, approximately matching the  $507{,}849$ parameters of the \latents model.
For the missing momentum regression task, \baselineL widens the first two DeepMET hidden dimensions from $(64, 32, 16)$ to $(128, 64, 16)$, raising the parameter count from $4{,}564$ to $13{,}044$, approximately matching the $12{,}756$ parameters of the \latents model.
The additional capacity is distributed across the hidden layers rather than concentrated in the output head so that the \baselineL model has more room to learn richer features rather than only a wider final projection, making the capacity-control comparison as generous as possible to the \baselineL model.
All other training hyperparameters are identical to those of the \baseline model.

\begin{figure}[htbp]
\centering
\includegraphics[width=0.40\textwidth]{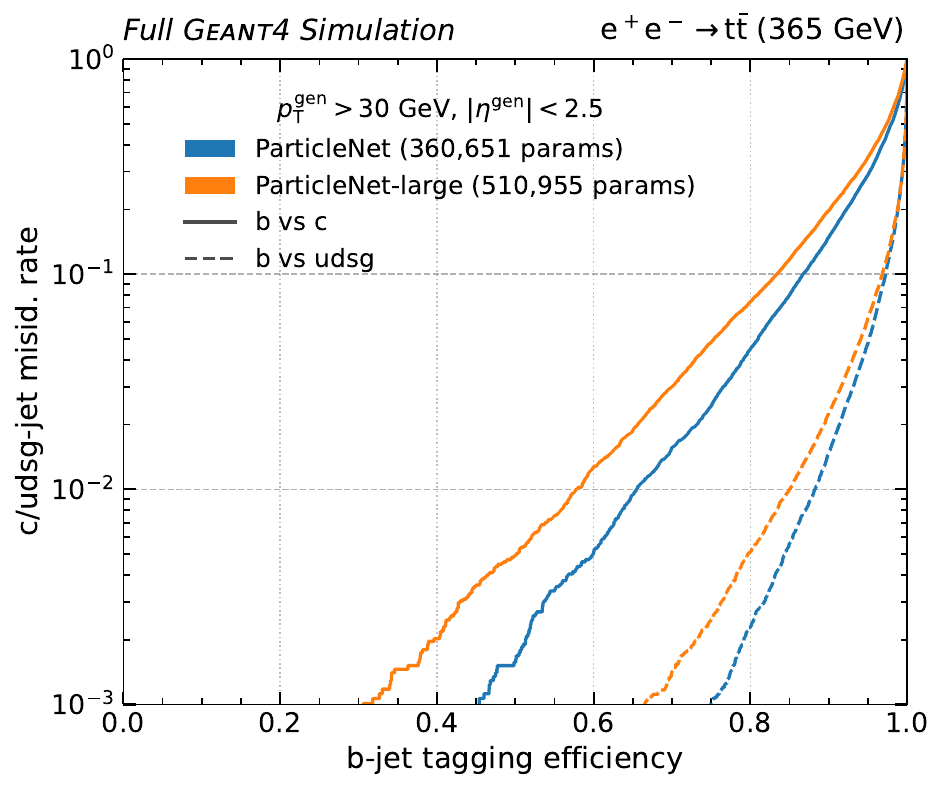}
\caption{
Capacity-control comparison for jet flavor identification: ROC curves at the full training set, evaluated on the test set, for the \baseline (blue) and \baselineL (orange) models.
Dashed lines show the light-flavor jet mis-identification rate vs \PQb-jet identification efficiency; solid lines show the \PQc-jet mis-identification rate.
A fiducial selection of $\pt^{\mathrm{gen}} > 30\GeV$ and $|\eta^{\mathrm{gen}}| < 2.5$ is applied.
The \baselineL model is trained with a single random seed, and \baseline shows its best seed (i.e., lowest validation loss).
}
\label{fig:app:btag}
\end{figure}

Across all three tasks the \baseline and \baselineL models perform comparably on their physics-performance metrics, with the \baselineL slightly trailing on jet flavor identification.
For jet flavor identification (Fig.~\ref{fig:app:btag}), the AUCs at the full training set are $0.990$ (\baseline) versus $0.988$ (\baselineL) against light-flavor jets, and $0.952$ versus $0.938$ against \PQc jets.
The \baselineL model is in fact worse than the \baseline model on both AUCs, marginally against light-flavor jets and more substantially against \PQc jets, confirming that the larger parameter count does not improve performance.

\begin{figure}[htbp]
\centering
\includegraphics[width=0.42\textwidth]{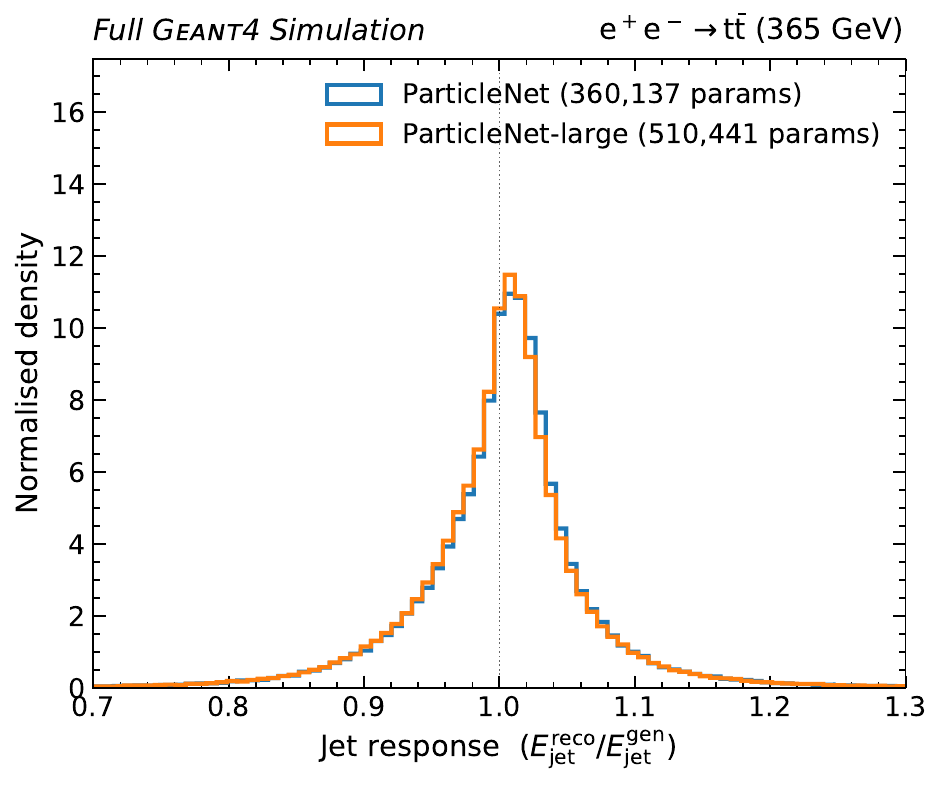}
\caption{
Capacity-control comparison for jet energy regression: inclusive jet-energy response distribution at the full training set, evaluated on the test set, for the \baseline (blue) and \baselineL (orange) models.
The \baselineL model is trained with a single random seed, and \baseline shows its best seed (i.e., lowest validation loss).
}
\label{fig:app:jet}
\end{figure}

For jet energy regression (Fig.~\ref{fig:app:jet}), the inclusive jet-energy response distributions agree to within $0.002$ in the median ($1.007$ for the \baseline model versus $1.005$ for \baselineL model), with both IQRs at approximately $0.057$.

\begin{figure}[!htbp]
\centering
\includegraphics[width=0.42\textwidth]{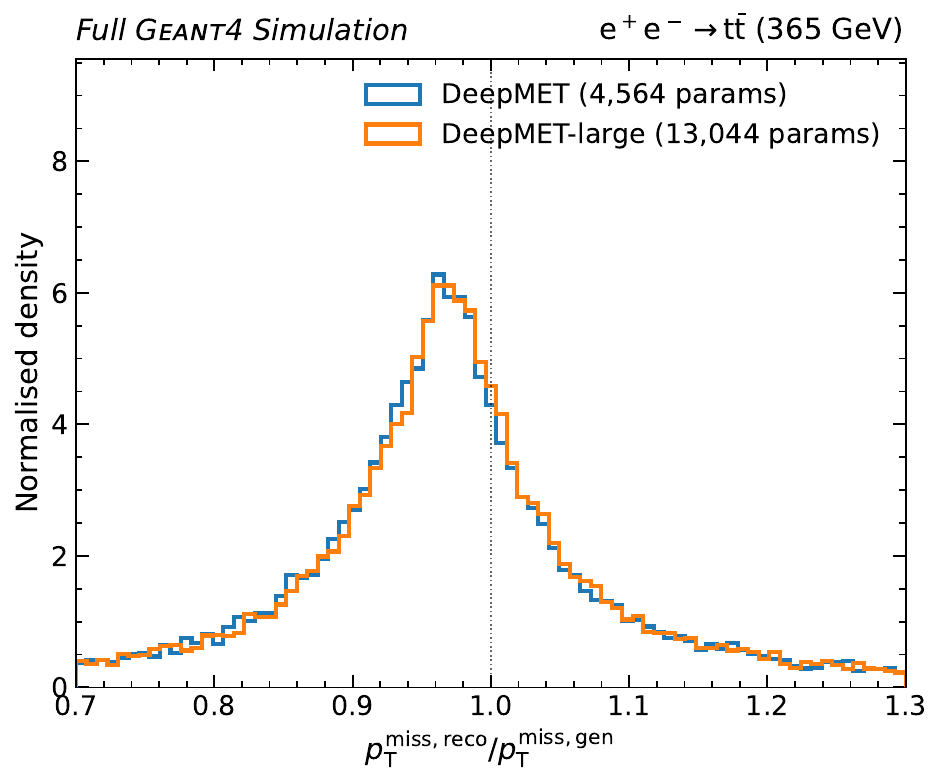}
\caption{
Capacity-control comparison for missing momentum regression: inclusive \ptmiss response distribution at the full training set, evaluated on the test set, for the \baseline (blue) and \baselineL (orange) models.
The \baselineL model is trained with a single random seed, and \baseline shows its best seed (i.e., lowest validation loss).
}
\label{fig:app:met}
\end{figure}

For missing momentum regression (Fig.~\ref{fig:app:met}), the inclusive \ptmiss response distributions are essentially indistinguishable, with medians of $0.967$ for the \baseline model and $0.970$ for the \baselineL model, and IQRs of approximately $0.123$ for both.
This confirms that the performance gains observed in the paper are not a consequence of the increased parameter count, but reflect genuine additional information supplied by the MLPF latent representations.

\section{Nonlinear probe of the latent representations}
\label{app:nlp}

The \probe model of Sec.~\ref{sec:setup:design} captures only the \emph{linearly} accessible information in the latent space, so its gap relative to the \baseline model could reflect either information absent from the latent representations or information present but not linearly separable.
To disentangle these, we train \probeL, the full task-specific architecture applied to the 128-dimensional MLPF latent representations alone, with all kinematic, PID, and track-displacement inputs removed.

The architecture is ParticleNet for jet flavor identification and jet energy regression, and DeepMET for missing momentum regression, identical to the \baseline models except that the input is the latent vector.
For the ParticleNet tasks, the graph building in the first layer is built from the constituent angular coordinates $(\eta, \phi)$ and the subsequent layers from the dynamically updated latent features, following the standard ParticleNet construction.
For DeepMET, the latent vector determines the per-candidate weights entering the \pvecmiss sum.
As for the other variants, we train five models with different random seeds and report the best model on the test set.

\begin{figure}[htbp]
  \centering
  \includegraphics[width=0.36\textwidth]{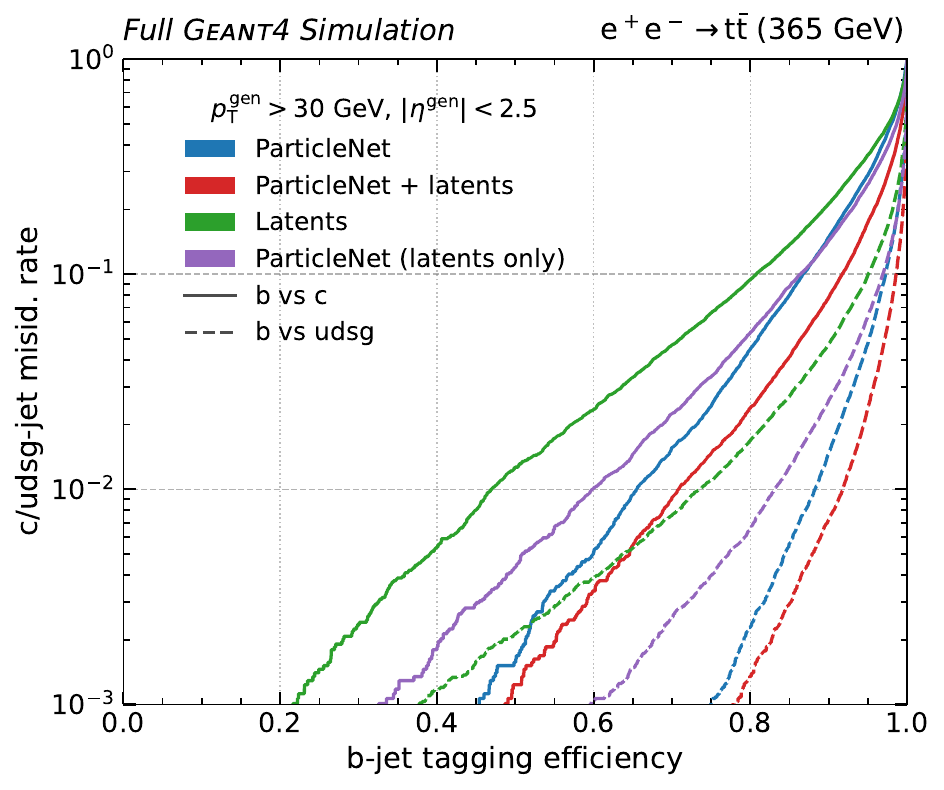}
  \caption{
  ROC curves for the \probeL (purple) model alongside the \baseline (blue), \latents (red), and \probe (green) models, trained on the full training dataset and evaluated on the test set, showing the mis-identification rate for light-flavor jets (dashed) and \PQc jets (solid) as a function of \PQb-jet identification efficiency, with a fiducial selection of $\pt^{\mathrm{gen}} > 30\GeV$ and $|\eta^{\mathrm{gen}}| < 2.5$.
  }
  \label{fig:app:nlp_btag}
\end{figure}

For jet flavor identification (Fig.~\ref{fig:app:nlp_btag}), at a $1\%$ mis-identification rate the \probeL model reaches a \PQb-jet identification efficiency of approximately $83\%$ against light-flavor jets and $60\%$ against \PQc jets, compared with $88\%$ and $66\%$ for the \baseline model and substantially exceeding the \probe model ($74\%$ and $47\%$).
The \probeL model thus recovers most of the \baseline discrimination using no kinematic or track-displacement features, while the \latents model remains the best-performing variant, indicating that the explicit track-displacement features still add complementary information beyond the latent space.

\begin{figure}[htbp]
  \centering
  \includegraphics[width=0.45\textwidth]{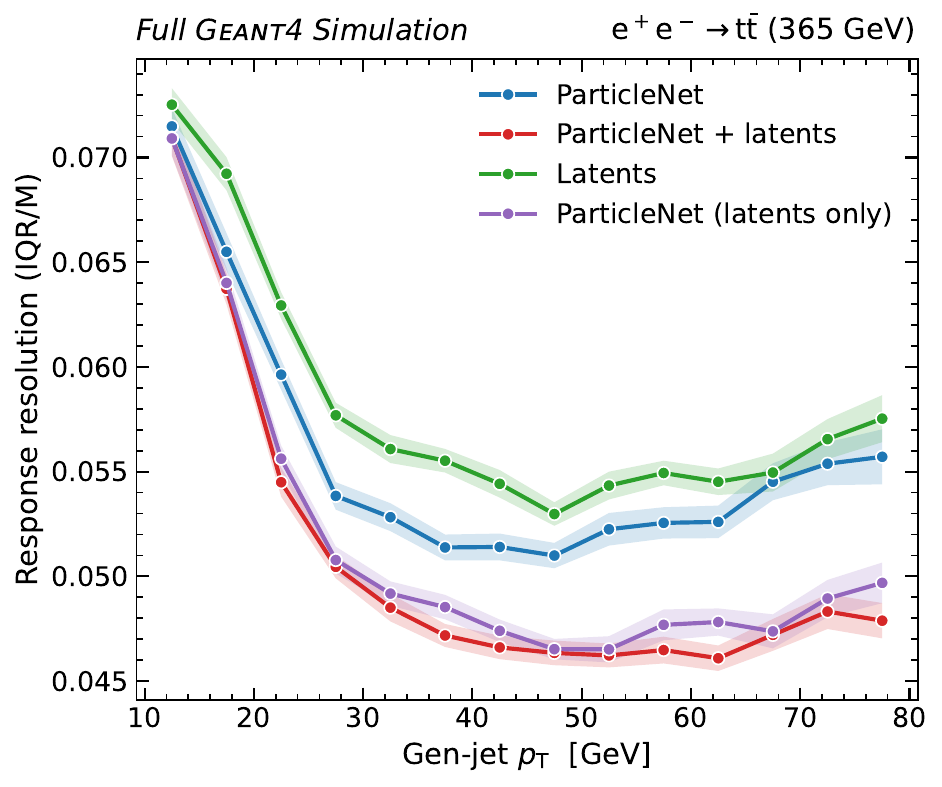}
  \caption{
  Jet energy resolution as a function of the generator-level jet \pt of the \probeL (purple) model alongside the \baseline (blue), \latents (red), and \probe (green) models at the full training set, evaluated on the test set.
  Shaded bands indicate the per-bin statistical uncertainty.
  }
  \label{fig:app:nlp_jer}
\end{figure}

For jet energy regression, the inclusive jet energy response has an IQR of $0.052$, matching the \latents model and improving on the \baseline model by approximately $10\%$, and the resolution is comparable to the \latents model across the full \pt range as shown in Fig.~\ref{fig:app:nlp_jer}.

\begin{figure}[htbp]
\centering
\includegraphics[width=0.42\textwidth]{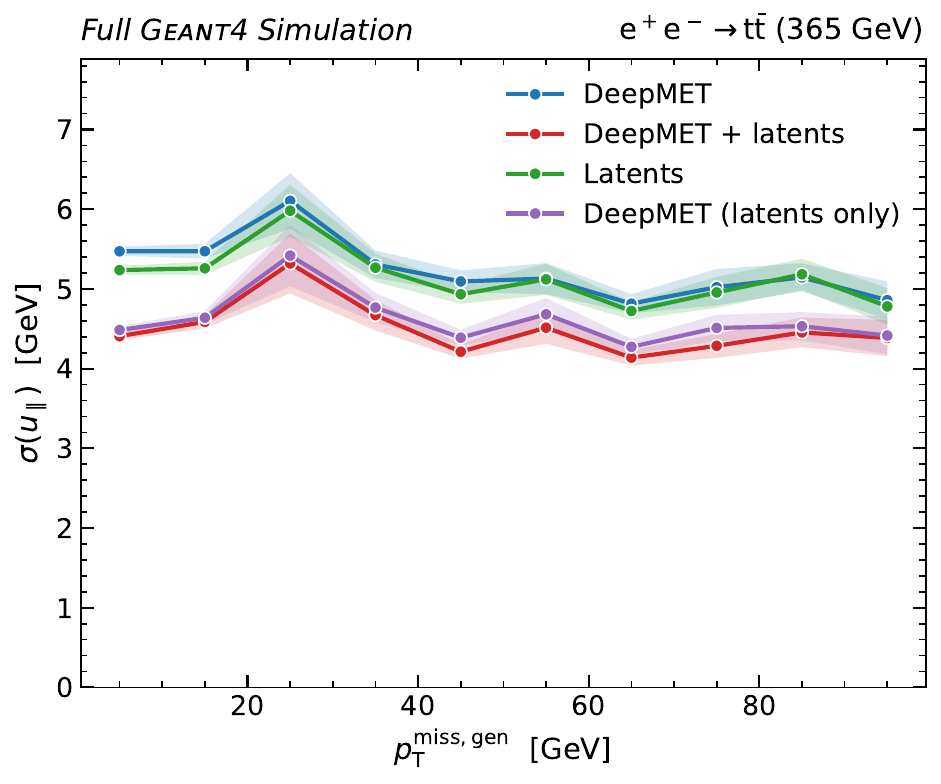}
\includegraphics[width=0.42\textwidth]{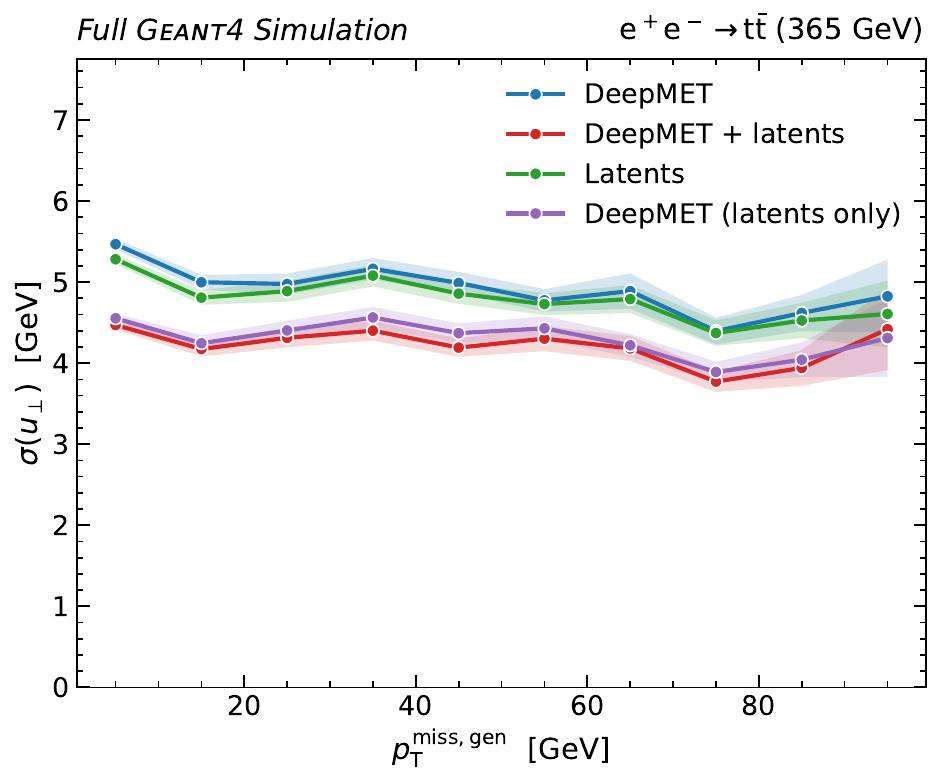}
\includegraphics[width=0.42\textwidth]{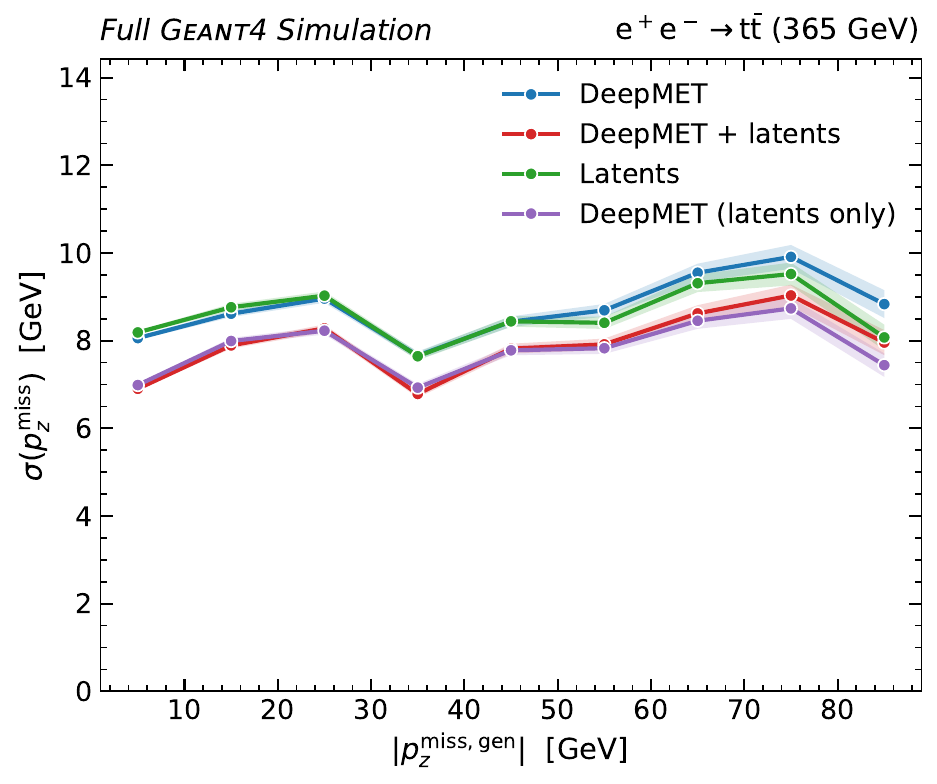}
\caption{
Missing momentum performance of the \probeL (purple) model alongside the \baseline (blue), \latents (red), and \probe (green) models at the full training set, evaluated on the test set.
Top, middle: recoil resolution parallel ($\sigma(u_\parallel)$) and perpendicular ($\sigma(u_\perp)$) to the generator-level \ptmiss direction, as a function of the \ptmiss magnitude.
Bottom: longitudinal missing-momentum resolution $\sigma(p_z^{\mathrm{miss}})$ as a function of $|p_z^{\mathrm{miss,gen}}|$.
Shaded bands indicate the per-bin statistical uncertainty.
}
\label{fig:app:nlp_met}
\end{figure}

For missing momentum regression (Fig.~\ref{fig:app:nlp_met}), the recoil resolution improves over the \baseline model by approximately $12\%$ in both projections and the longitudinal resolution by approximately $11\%$, comparable to the \latents model.

Across the three tasks, a model trained on the latent representations alone matches the \latents model for the two regression tasks, indicating that the latent space captures essentially all of the task-relevant information there.
For jet flavor identification, it approaches but does not reach the \baseline model, because the \baseline model additionally exploits explicit track-displacement variables---the IP significances and the signed 2D and 3D IPs---that are constructed downstream from the track covariance and the jet axis and are therefore not encoded in the latent representations.
Even without these inputs, the \probeL model recovers most of the \baseline model discrimination, confirming that flavor-relevant information is already accessible to some extent from the latent space.

\clearpage

\bibliographystyle{cms_unsrt}
\bibliography{references}

\end{document}